\documentclass[oneside,english,latin9,usenames,table,dvipsnames,svgnames]{aa}
\usepackage[T1]{fontenc}
\usepackage{geometry}
\usepackage{color}
\usepackage{babel}
\usepackage{textcomp}
\usepackage{url}
\usepackage{amsmath}
\usepackage{graphicx}
\usepackage[unicode=true,pdfusetitle,
 bookmarks=true,bookmarksnumbered=false,bookmarksopen=false,
 breaklinks=false,pdfborder={0 0 0},pdfborderstyle={},backref=false,colorlinks=true]
 {hyperref}
\hypersetup{
 urlcolor=blue,citecolor=blue}

\makeatletter

\providecommand{\tabularnewline}{\\}

\newcommand{\eROSITA}{\textit{eROSITA}}

\newcommand{\threshworstAMPLSIGbandsoft}{2.6}

\newcommand{\falseposnumAMPLSIGbandsoft}{0.1}

\newcommand{\threshworstNEVSIGbandsoft}{1.7}

\newcommand{\falseposnumNEVSIGbandsoft}{0.1}

\newcommand{\threshworstFVARSIGbandsoft}{3.3}

\newcommand{\falseposnumFVARSIGbandsoft}{0.1}

\newcommand{\threshworstSCATTLObandsoft}{0.14}

\newcommand{\falseposnumSCATTLObandsoft}{13.3}

\makeatother

\begin{document}
\title{Systematic evaluation of variability detection methods for eROSITA}
\abstract{The reliability of detecting source variability in sparsely and irregularly
sampled X-ray light curves is investigated. This is motivated by the
unprecedented survey capabilities of eROSITA onboard SRG, providing
light curves for many thousand sources in its final-depth equatorial
deep field survey. Four methods for detecting variability are evaluated:
excess variance, amplitude maximum deviations, Bayesian blocks and
a new Bayesian formulation of the excess variance. We judge the false
detection rate of variability based on simulated Poisson light curves
of constant sources, and calibrate significance thresholds. Simulations
with flares injected favour the amplitude maximum deviation as most
sensitive at low false detections. Simulations with white and red
stochastic source variability favour Bayesian methods. The results
are applicable also for the million sources expected in eROSITA's
all-sky survey.}
\author{Johannes Buchner\inst{1}\thanks{\protect\href{mailto:mailto:johannes.buchner.acad@gmx.com}{johannes.buchner.acad@gmx.com}}\and Thomas
Boller\inst{1}\and David Bogensberger\inst{1}\and Adam Malyali\inst{1}\and Kirpal
Nandra\inst{1}\and Joern Wilms\inst{2}\and Tom Dwelly\inst{1}\and Teng
Liu\inst{1}}
\institute{Max Planck Institute for Extraterrestrial Physics, Giessenbachstrasse,
85741 Garching, Germany\and Dr. Karl Remeis-Observatory and Erlangen
Centre for Astroparticle Physics, Friedrich-Alexander-Universität
Erlangen-Nürnberg,Sternwartstr. 7, 96049 Bamberg, Germany}
\date{-Received date / Accepted date}
\titlerunning{X-ray Variability Methods}
\authorrunning{Buchner et al.}
\maketitle

\section{Introduction}

The variability of astrophysical sources is a powerful diagnostic
to differentiate between different physical models even when those
models predict similar spectral energy distributions. Variability
studies have enriched the zoo of astrophysical phenomena with new
mysteries, including in recent years for example fast radio bursts
\citep{Lorimer2007,Petroff2019}, ultra-luminous X-ray sources \citep[e.g.,][]{Bachetti2014,Liu2013}
and quasi-periodic eruptions \citep{Miniutti2019}. In high-energy
astrophysics, the search for transient phenomena has a long history
with gamma-ray bursts \citep{Klebesadel1973,Gehrels2012}, for example.
Missions such as MAXI \citep{2009PASJ...61..999M}, Rossi X-ray Timing
Explorer \citep{Swank2006} and Swift \citep{Gehrels2004} were explicitly
designed to characterize the variable X-ray sky. However, these missions
are sensitive only to the brightest objects (typically fewer than
one hundred variability triggers per year). This situation has changed
with the launch of the \emph{eROSITA} telescope on-board \emph{SRG}
\citep{Predehl2020}, and its all-sky monitoring every 6 months in
the first 4 years. Because \emph{eROSITA} scans the X-ray sky rapidly
over large areas down to faint fluxes, it has the potential to reveal
a myriad of diverse variable and transient phenomena. Preliminary
analysis of the first, most extreme events revealed gamma-ray burst
afterglows \citep{2020GCN.26988....1W}, super-soft emission from
a classical nova \citep{Ducci2020}, flares in millisecond pulsars
\citep{Koenig2020}, flares of unknown origin \citep{Wilms2020} and
new types of tidal disruption events \citep{Malyali2021}. These phenomena
exhibit different variability behaviour (e.g., flares or red noise).
To fully exploit the \emph{eROSITA} dataset we require robust and
well characterized techniques to identify, classify and characterize
the variability properties of each detected X-ray source.

Identifying source variability in the X-rays is no small task. In
the recent, large-scale optical photometric surveys (Gaia, Zwicky
Transient Factory, Optical Gravitational Lensing Experiment, etc.),
systematics typically dominate measurement uncertainties, requiring
a machine learning classifier to postprocess various classical light
curve summary statistics \citep{Debosscher2007,Kim2011,Palaversa2013,Masci2014,Armstrong2016,Holl2018,Heinze2018,Jayasinghe2019,vanRoestel2021}.
In contrast, for repeated X-ray surveys where most X-ray sources are
found near the detection limit, statistical (Poisson) uncertainties
are dominant. In this regime, methods such as fractional variance
\citep{Edelson1990}, excess variance \citep{Nandra1997} and Bayesian
blocks \citep{Scargle2013} have been proposed. However, their application
has typically been limited to a handful of light curves at a time.
\citet{DeLuca2021} investigated variable objects of the archival
\emph{XMM-Newton} X-ray sky with Bayesian blocks and light curve summary
statistics\textbf{.} \emph{eROSITA} detected almost a million point
sources already in its first all-sky survey\textbf{ }(eRASS1), and
a new all-sky survey with similar characteristics is conducted every
0.5 years. With this many sources, the calibration of the instrument
and detection methods becomes important to avoiding both false positives
and false negatives in large numbers. Additionally, \emph{eROSITA's}
scanning pattern imprints strong temporal modulations of the effective
instrument sensitivity at any particular sky location. To summarise,
any useful method must consistently distinguish Poisson and sensitivity
fluctuations from variations intrinsic to the astrophysical source.
For these reasons, we have examined the performance of commonly used
variability analysis methods, together with a novel Bayesian approach,
within the eROSITA regime. 

This paper investigates the reliability and sensitivity of various
variability detection methods, based on a pilot \emph{eROSITA} survey
over a extra-galactic 140 square degrees field. Its characteristics,
such as exposure depth, are similar to the final stacked eight-year
all-sky surveys. Classes of variable sources expected include flaring
X-ray stars and variable active galactic nuclei. The properties of
that data set is presented in §\ref{sec:Data}, including how counts
are extracted in time bins. Various ways for constructing light curves
(§\ref{subsec:perbinmethods}) and visualising them (§\ref{subsec:viz})
are discussed. These light curves form the foundation for the considered
variability detection methods, which are presented in detail in §\ref{subsec:Methods-for-variability}.
Section~\ref{subsec:Method-comparison} explains our methodology
for evaluate and compare the methods, based on extensive numerical
simulations (§\ref{subsec:Simulation-setup}). The results section
presents the calibrations needed for reliable use of the methods (§\ref{subsec:Thresholds-for-low})
and how sensitive they are to various types of variability (§\ref{subsec:Sensitivity-evaluation}).
We conclude with discussing in §\ref{sec:Discussion-and-Conclusion}
the advantages of a method newly developed in this work, Bayesian
excess variance, and its future use for \emph{eROSITA} and beyond.

\section{Data\label{sec:Data}}

\begin{figure}
\includegraphics[width=1\columnwidth]{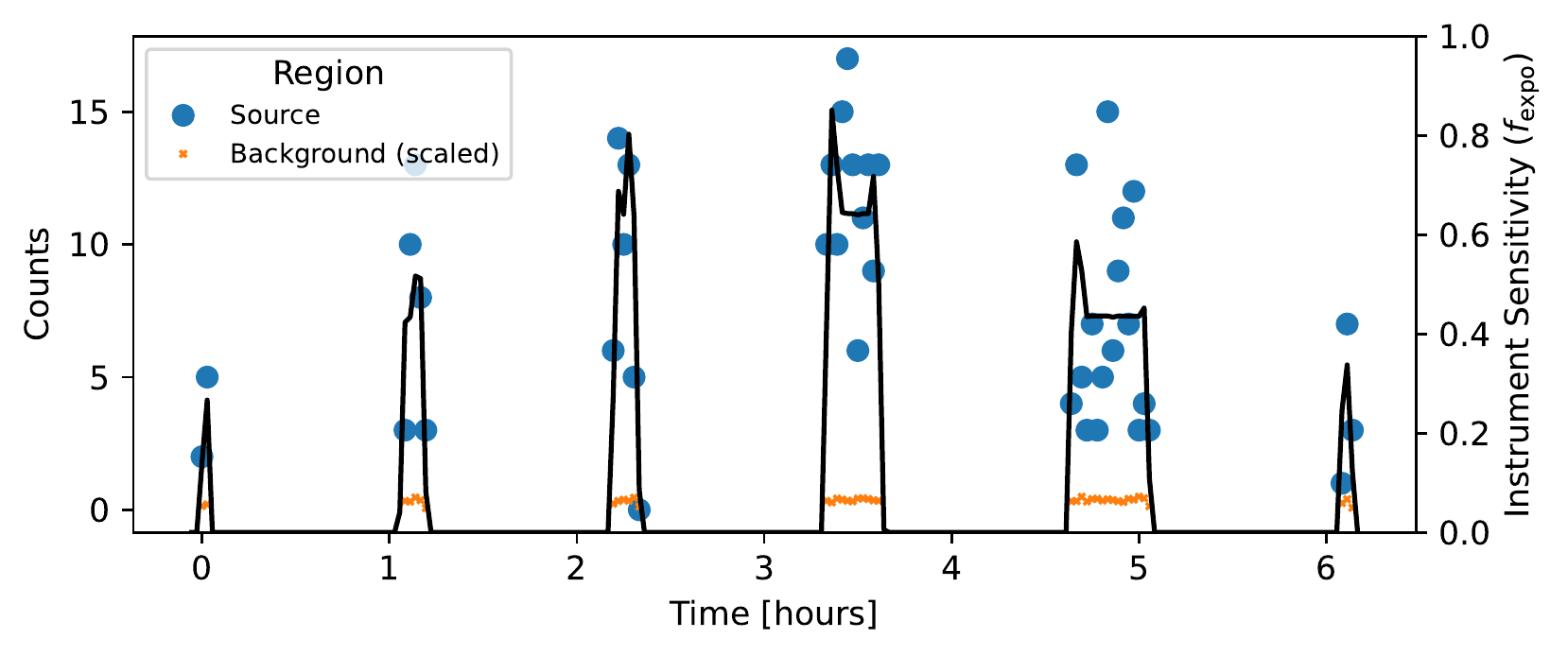}

\caption{\label{fig:rawdata}Example counts of the 10th brightest eFEDS source.
The markers indicate the number of counts from a single \emph{eROSITA}
telescope module in 100s time bins, recorded over a six hour period.
The black curve shows the sensitivity to the source position over
time.}
\end{figure}

The eROSITA Final Equatorial-Depth Survey (eFEDS) field was observed
with \eROSITA{} in November 2019. The source catalog paper \citep{Brunner2021}
presents the observations, eROSITA analysis software and data treatment.
Survey aspects that are important for investigating source variability
are highlighted in this section. The depth expected after completion
of all eROSITA all-sky survey scans was reached and slightly exceeded
in the eFEDS field. eFEDS consists of four adjacent, approximately
rectangular areas aligned with the Ecliptic coordinate grid, which
were covered from Ecliptic east to west by a sequence of linear scans
going from Ecliptic north to south and back. The typical scanning
speed was \textbf{$13.15\,''/\mathrm{s}$} \citep{Brunner2021}. Because
the field of view (FoV) of \eROSITA{} is about ten time larger than
the distance between scans, each source was covered multiple times.
The resulting cadence is such that sources were visible continuously
for several minutes, and revisited approximately every hour. The black
curve of Figure~\ref{fig:rawdata} illustrates this strongly variable
instrument sensitivity over time for a typical source. This illustrates
the difference between eROSITA survey light curves and those of typical
pointed observations, where the instrument sensitivity is nearly constant.
Therefore different analysis methods are required. In eFEDS, 27910
point sources were detected \citep{Brunner2021} in the 0.2-2.3~keV
band. These form the main eFEDS sample, which is also the basis of
this paper.

For all 27910 sources, a spectrum and light curve was extracted. The
procedure is described in detail in \citet{2021arXiv210614522L}.
Source counts are extracted from a circular aperture of radius $\approx20-40''$
(increasing with source counts). A representative local background
is extracted from an annular region, also centred at the source position.
The inner and outer radii of the annulus are scaled to be 5 and 25
times larger than the source radius, which yields a background to
source area ratio of $r'\approx200$. Because most \eROSITA{} observations
are made in scanning mode, these extraction regions are defined in
terms of sky coordinates (rather than on some instrumental coordinate
system). Neighbouring sources are masked from the source and background
regions before extraction, and finally the light curves from the seven
telescope modules are summed \citep[see][]{2021arXiv210614522L}.
Figure~\ref{fig:rawdata} shows an example of counts extracted over
time for a bright source, with the background counts scaled according
to the area ratio. The count statistics are low and therefore in the
Poisson regime.

\begin{figure}
\includegraphics[width=1\columnwidth]{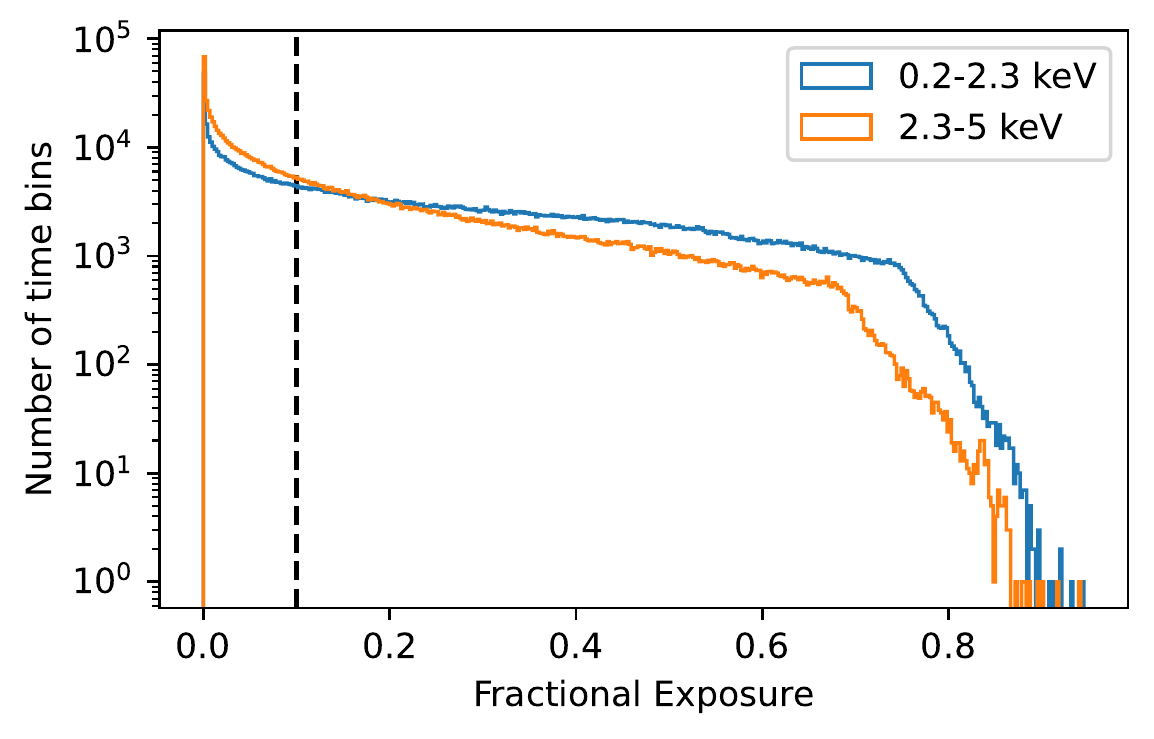}
\caption{Distribution of fractional exposure ($f_{\mathrm{expo}}$) values
for all light curve time bins.}
\label{fig:fieldstats-fracexp}
\end{figure}

Light curves are extracted in three bands. Their energy ranges are
0.2-5 (full band, band 0), 0.2-2.3 (soft band, band 1), and 2.3-5
keV (hard band, band 2). Here, we focus primarily on the soft, and
secondarily on the properties of the hard band light curves. Because
of \eROSITA{}'s relatively soft X-ray response, the full band is
dominated by, and nearly identical to, the soft band for most detected
sources.

Light curves with time bins of 100s were constructed using \texttt{srctool}\footnote{\url{https://erosita.mpe.mpg.de/eROdoc/tasks/srctool\_doc.html}}
\citep[version eSASSusers\_201009]{Brunner2021}. This binning choice
balances samples times during the survey track as well as revisits.
The effective sensitivity of \eROSITA{} to an astrophysical source
varies with time, as the source moves through the FoV of the telescope
modules, and becomes zero in time bins while the source is outside
the FoV. The dimensionless fractional exposure ($f_{\mathrm{expo}}$)
parameter (range 0 to 1) computed by \texttt{srctool,} is an estimate
of the effective sensitivity of the instrument within a time bin to
the source in question. The computation of $f_{\mathrm{expo}}$for
each bin of the light curve is carried out by integrating the instantaneous
effective response of the instrument within the bin, on a time grid
comparable to the instrumental integration time (Delta t=50 ms). The
computation of $f_{\mathrm{expo}}$takes into account the geometry
of the source extraction aperture, the telescope attitude, off-axis
vignetting, energy- and position-dependent point spread function (PSF),
good time intervals, instrument dead time, and the location any bad
pixels, and is carried out independently for each of the seven telescope
modules. An effective spectral index of Gamma = 1.7 is assumed when
weighting the energy-dependent components of the instrument response
model (vignetting, PSF) across broad energy bins. We normalize $f_{\mathrm{expo}}$
relative to the response expected for an on-axis point source observed
with all telescope modules, assuming no extraction aperture losses.
The black curve of Figure~\ref{fig:rawdata} gives an example of
the $f_{\mathrm{expo}}$ windowing for an arbitrarily chosen source
in eFEDS, decreasing when the source is at the border of the field
of view. Figure~\ref{fig:fieldstats-fracexp} shows the $f_{\mathrm{expo}}$
of all time bins for sources in the eFEDS field. Because the current
understanding of the \eROSITA{} vignetting function is somewhat
uncertain at large off-axis angles, we only consider time bins exposed
to $f_{\mathrm{expo}}>0.1$. This cut tends to segment the light curves
into disjoint intervals that are filled with meaningful data. As the
reflectivity of the eROSITA optics reduces at higher energies and
large grazing angles, the hard band has a smaller effective field
of view and systematically lower $f_{\mathrm{expo}}$ values (see
Figure~\ref{fig:fieldstats-fracexp} ). Unless otherwise stated,
the remainder of this paper assumes that $f_{\mathrm{expo}}$ represents
the relative sensitivity of the instrument correctly.

\begin{figure}
\includegraphics[width=1\columnwidth]{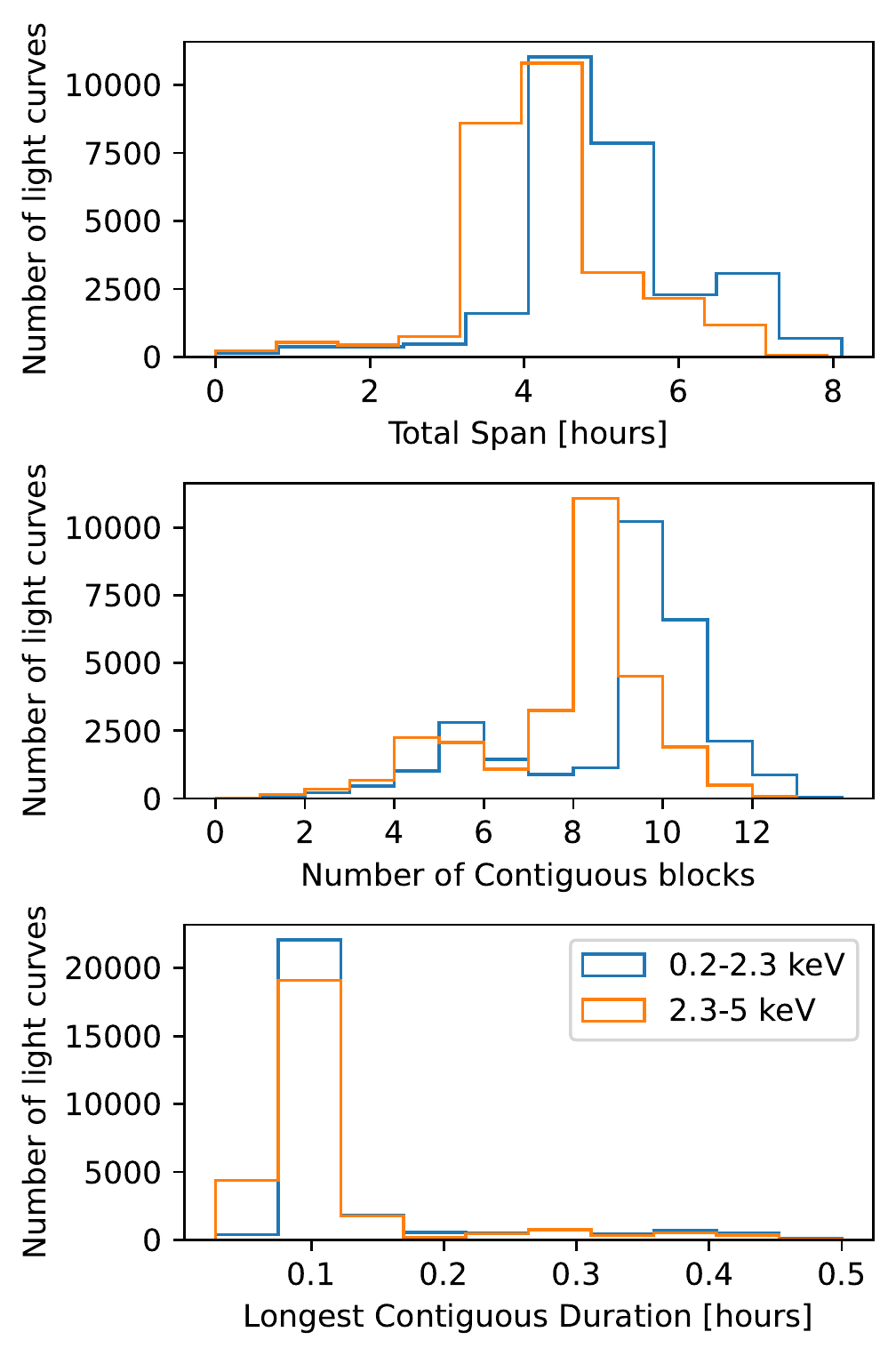} \caption{Light curve cadence summary statistics. The \emph{top panel} shows
the time between first and last exposed time bins for each light curve,
with typical values of four to seven hours. As Figure~\ref{fig:rawdata}
illustrates, the lightcurves are segmented into blocks. The \emph{middle
panel} counts the number of blocks, which range from four to twelve.
The\emph{ bottom panel }shows the duration of the longest block for
each light curve, which is typically last only a few minutes.}
\label{fig:fieldstats-t}
\end{figure}

The cadences effectively sampled by the light curves are summarized
in Figure~\ref{fig:fieldstats-t}. The top panel illustrates that
sources are typically observed over a span of four to seven hours.
During this time, the light curves exhibit several gaps (see Figure~\ref{fig:rawdata}),
during which other parts of the field were scanned. Typically, there
are four to twelve blocks of contiguous observations (middle panel
of Figure~\ref{fig:fieldstats-t}), lasting not more than a few minutes
each. This results in two effective cadences: consecutive exposures
lasting a few minutes and re-visits on hour time-scales.

\begin{figure}
\includegraphics[width=1\columnwidth]{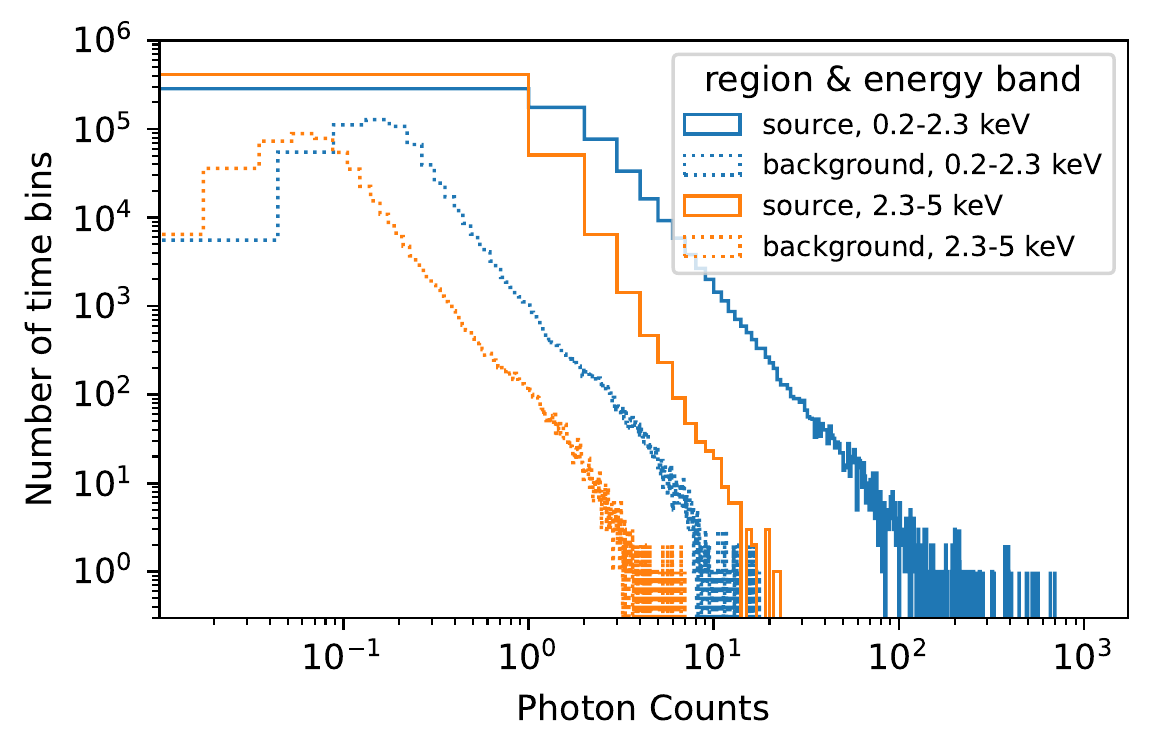}
\caption{Source and background photon count distribution in the soft and hard
bands. The background counts are scaled by the source to background
area ratio.}
\label{fig:fieldstats-counts}
\end{figure}

For each time bin the observed source counts $S$ and background counts
$B$ are listed. Figure~\ref{fig:fieldstats-counts} presents histograms
of these counts, for the soft (blue) and hard (orange) band. For most
time bins, the number of counts is in the single digits, contributed
by sources near the detection threshold. \eROSITA{} is most sensitive
in the soft band, which typically shows more counts (up to 100~cts/bin),
approximately 20 times higher than the maximum seen in the hard band.
The expected number of background counts in each time bin is typically
below 1. Light curves are presented in \citet{2021arXiv210614523B},
and significantly variable sources are identified. This focus of this
work is to investigate methods to determine whether sources are significantly
variable.

\section{Methods}

The counts observed in a time bin $t$ can be expressed as a Poisson
process, which integrates the band\emph{ }count rate $R$, dampened
by the efficiency $f_{\mathrm{expo}}$ within the time interval $\Delta t$.
For the background region and assuming $f_{\mathrm{expo}}$ is constant
within the time bin, this can be written as

\begin{equation}
B\sim\mathrm{Poisson}\left(R_{B}\times f_{\mathrm{expo}}\times\Delta t\right)\label{eq:bkgpoisson}
\end{equation}
where $C\sim\mathrm{Poisson}(\lambda)$ means ``Counts $C$ is a
Poisson random variable with a mean of $\lambda$''. The total counts
in the source region, $S$, contain contributions from the source,
with count rate $R_{S}$ and background: 
\begin{equation}
S\sim\mathrm{Poisson}\left(\left(R_{S}+R_{B}\times r\right)\times f_{\mathrm{expo}}\times\Delta t\right)\label{eq:srcpoisson}
\end{equation}
The background rate $R_{B}$ is scaled by the area ratio of source
and background extraction regions, $r$, with values near 1\% being
typical. The unknowns are $R_{B}$, the background count rate and
$R_{S}$, the net (without background) source count rate. Typical
values for $R_{S}/(R_{B}\times r)$ are 8 for the soft band and 1
for the hard band.

\subsection{Methods to infer the source count rate per-bin\label{subsec:perbinmethods}}

In the following, we present two approaches for inferring the net
count rate $R_{S}$ in each time bin.

\subsubsection{Classic per-bin source rate estimates\label{sec:srcestimates}}

The classic point estimator for the net source count rate $R_{S}$
is: 
\begin{equation}
\hat{R}_{S}=\frac{S-B\times r}{f_{\mathrm{expo}}\times\Delta t}\label{eq:srcrate}
\end{equation}
Here, the background count rate in the background region is estimated
with: 
\begin{equation}
\hat{R_{B}}=\frac{B}{f_{\mathrm{expo}}\times\Delta t}\label{eq:bkgrate}
\end{equation}
The uncertainty in the net source count rate $\hat{R}_{S}(t)$ is
estimated as: 
\begin{equation}
\hat{\sigma}(\mathrm{R}_{S})(t)=\frac{\sqrt{\hat{\sigma}(S)(t)^{2}+\hat{\sigma}(B)(t)^{2}\times r}}{f_{\mathrm{expo}}(t)\times\Delta t}\label{eq:srcrateerr}
\end{equation}
Here, $\hat{\sigma}(C)$ (with $C$ either $S$ or $B$) is the uncertainty
of the expected number of counts, given the observed counts $C$.
One possibility is to use the simple $\hat{\sigma}(C)=\sqrt{C}$ estimator.
However, in the low count regime, the uncertainties are then severely
under-estimated (for example, when $S=0$). These leads to a strong
count-dependent behaviour of any method that ingests these uncertainties.

Confidence intervals for the Poisson processes have been studied extensively
in the X-ray and gamma-ray astronomy literature \citep[e.g.,][]{Gehrels1986,Kraft1991}.
They are asymmetric in general for realistic settings, and thus cannot
be readily propagated in equation~\ref{eq:srcrateerr}. We adopt
the upper confidence interval formula $\hat{\sigma}(C)=\sqrt{C+0.75}+1$
from \citep{Gehrels1986}, and use it also as a lower confidence interval,
instead of the formula $\hat{\sigma}(C)=\sqrt{C-0.25}$. This conservative
choice tends to enlarge the error bars, and thus makes the data appear
less powerful than they actually are. An alternative that is being
considered for future releases of srctool are maximum likelihood-derived
confidence intervals found by numerically exploring profile likelihoods
\citep{Barlow2003}.

\subsubsection{Bayesian per-bin source rate estimates\label{sec:srcestimatesbayes}}

A drawback of the estimates above is the Gaussianity error propagation.
In the low-count regime, the Poisson uncertainties become asymmetric.
We adopt the approach of \citet{Knoetig2014} to propagate the uncertainties
in a Bayesian framework.

Firstly, the unknown background count rate $R_{B}(t)$ only depends
on known quantities in eq.~\ref{eq:bkgpoisson}. The Poisson process
likelihood, $\mathrm{Poisson(C|\lambda)=\lambda^{k}\times e^{-\lambda}/k!}$,
can be combined with a flat, improper prior on the expected count
rate $\lambda=R_{B}\times f_{\mathrm{expo}}$, to define a posterior
that can be numerically inverted using the inverse incomplete Gamma
function $\Gamma^{-1}$ \citep[see also][]{Cameron2011}. Specifically,
the $q$-th quantile of the posterior probability distribution of
$R_{B}(t)$ can be derived as $\Gamma^{-1}(B+1,q)\times r/f_{\mathrm{expo}}$.
The same approach cannot be applied to $R_{S}$, because it depends
on the values of $S$ and $R_{B}(t)$ (eq.~\ref{eq:srcpoisson}).
We thus compute the marginalised likelihood function of the source
rate $R_{S}$ as: 
\begin{flalign}
P(S|R_{S}) & =\int_{R_{B}}P(B|R_{B})\times P(S|R_{S}+R_{B})\times\pi(R_{B})\label{eq:pobjnumerical}\\
 & =\int_{0}^{1}P(S|R_{S}+\Gamma^{-1}(B+1,q)\times r)\,dq\label{eq:pobjnumericalinv}
\end{flalign}
In practice, equation~\ref{eq:pobjnumerical} is evaluated by numerically
integrating over $q$ in a grid \citep[see][for an alternative method]{Knoetig2014}.

The goal is then to place constraints on $R_{S}$ using the likelihood
function $P(S|R_{S})$. Evaluating a grid over $R_{S}(t)$ over a
reasonable range (logarithmically between 0.01 and 100~cts/s) explores
the likelihood function of eq.~\ref{eq:pobjnumerical}. If the grid
points are interpreted to be equally probable a priori, quantiles
(median, $1\sigma$-equivalents) can then be read off the normalised
cumulative of $P(S|R_{S})$ grid values, and form Bayesian alternatives
for $\hat{R_{S}}(t)$ and $\hat{\sigma}$. A different approach to
the priors on $R_{S}$ is explored below in section \ref{subsec:Bayesian-excess-variance}.

\subsection{Visualisations\label{subsec:viz}}

\begin{figure}
\includegraphics[width=1\columnwidth]{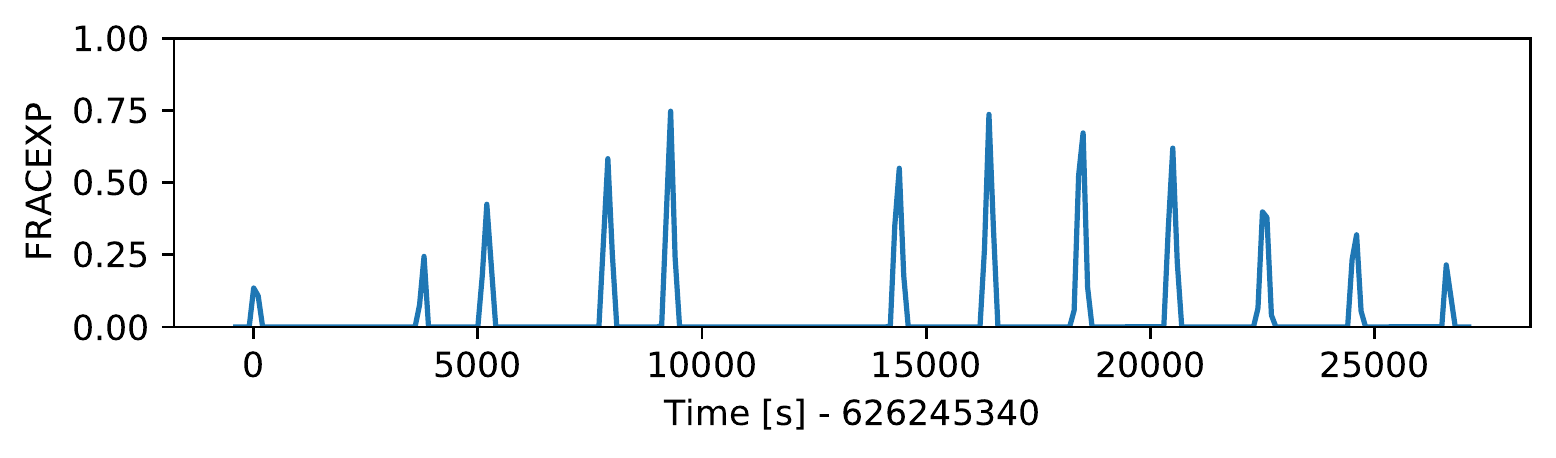}

\includegraphics[width=1\columnwidth]{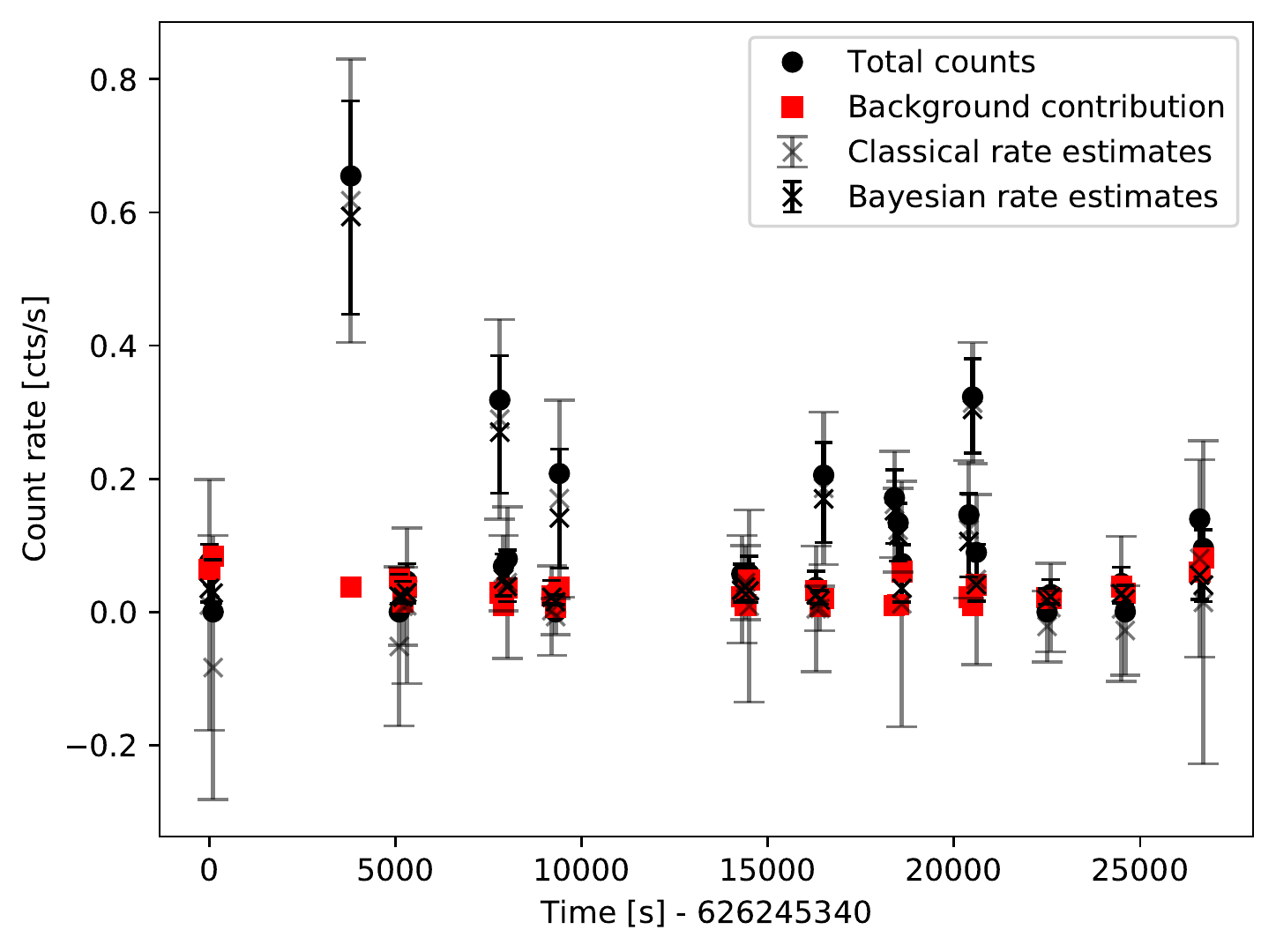}

\caption{\label{fig:viz}Visualisations. \emph{Top panel}: The fractional exposure
over time for an example source with 10 passes. \emph{Bottom panel}:
Light curve of a simulated source with constant count rate and a bright
flare. Black circles show the total counts without background subtraction,
red points show the expected background count rates in the source
region (eq.~\ref{eq:bkgrate}). Gray error bars show Classical net
source count rate estimates (eq.~\ref{eq:srcrate} and \ref{eq:srcrateerr}).
Black error bars show Bayesian net source count rate posterior distributions,
represented visually with 10\%, 50\% and 90\% quantiles under a log-uniform
prior (eq.~\ref{eq:pobjnumericalinv}).}

\end{figure}

The per-bin estimates defined above provide the possibility to plot
time series of inferred source count rates (a light curve). An example
(simulated) light curve is shown in Figure~\ref{fig:viz}, with both
Classical and Bayesian error bar estimates. The classical confidence
intervals are symmetric and sometimes include negative count rates.
The Bayesian estimates are asymmetric and always positive.

Some forms of variability can then be judged by identifying if the
count rates are consistent over time. In Figure~\ref{fig:viz}, one
may identify a major flare near t = 4000s, and perhaps two minor ones.
One shortcoming of this approach is that the judgment by eye is subjective
and difficult to reproduce. The Poisson fluctuations are also unintuitive
(there was only one real flare injected in this simulated time series).
Nevertheless, it can be insightful to try to understand what various
methods ``see'', and try to understand what likely triggered a statistical
test. They are also useful for judging the plausibility of the data
under current calibration. For example, if the fractional exposure
$f_{\mathrm{expo}}$ is mis-estimated at large off-axis angles, the
count rates are enhanced or reduced while the source enters and leaves
the field-of-view. Such systematic over- or under-corrections can
become visible as U or inverse U shaped light curves.

\begin{figure}
\begin{centering}
\includegraphics[width=1\columnwidth]{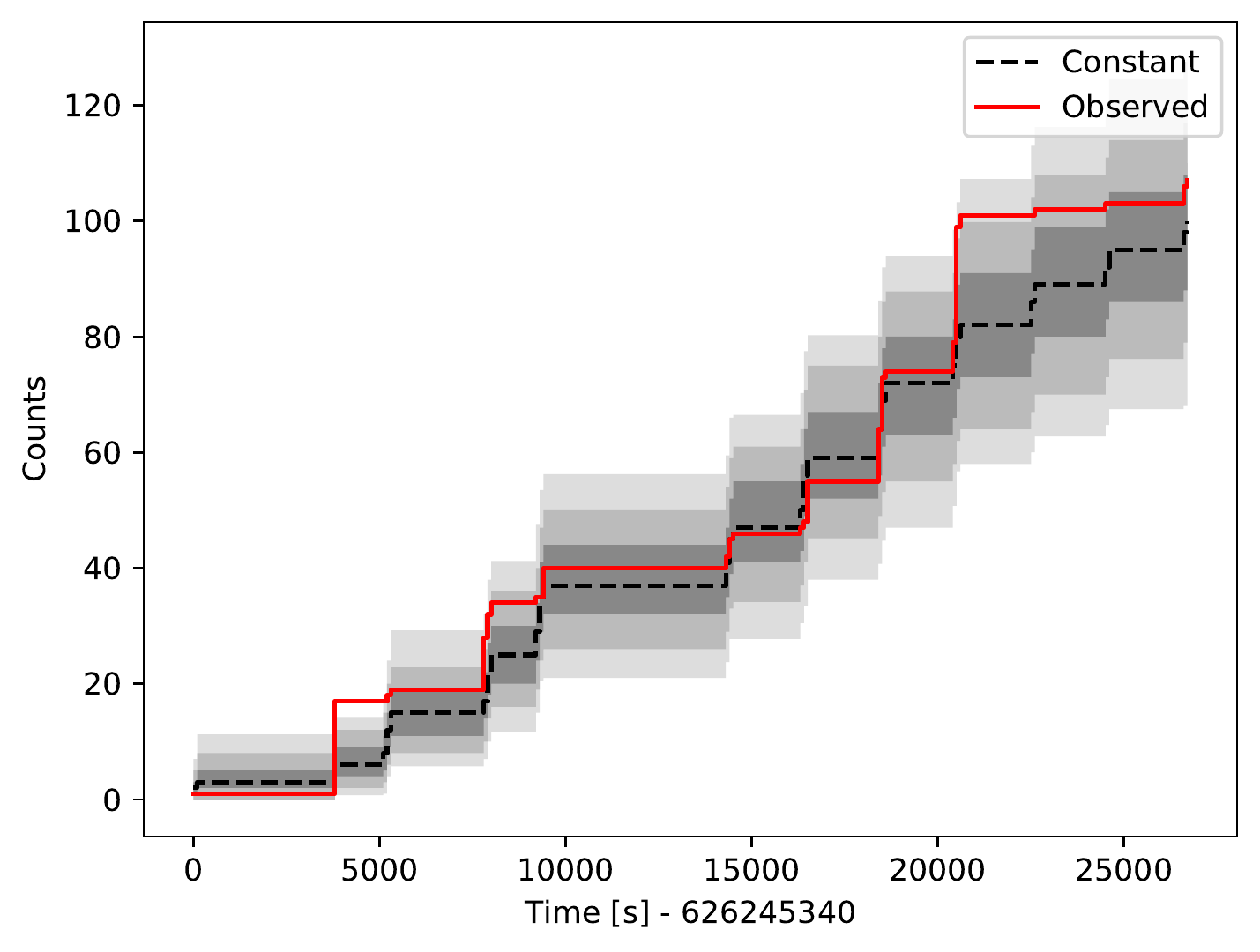}
\par\end{centering}
\caption{\label{fig:cumviz}Cumulative count visualisation. The red curve are
the observed cumulative counts over time. The expectation of a constant
source is indicated as a dashed curve and gray intervals corresponding
to 1, 2 and 3$\sigma$.}
\end{figure}

The binning of the time series also influences the visualisation.
Large time bins may average out short-term variability. Small time
bins may contain too few counts and thus large uncertainties. The
accumulation of nearby data points is difficult to do by eye. However,
this can be important, as variations are typically correlated on short
time-scales.

Cumulative counts address some of these limitations. Figure~\ref{fig:cumviz}
plots the cumulative counts in the source region, $S$, over time,
as a red curve. In the last time bin, all counts are noted. For judging
whether this light curve is variable, we generate Poisson counts for
1000 simulated time series assuming a constant source, following eq.~\ref{eq:srcpoisson}.
For this, the classically inferred source and background count rates
are assumed. The generated counts are shown in Figure~\ref{fig:cumviz},
with the mean as dashed black curve and intervals corresponding to
1, 2 and 3$\sigma$ as gray shadings. For this source, we see that
near 4000s, the red curve departs above the 3$\sigma$ range, indicating
an excess of counts at that time (consistent with \ref{fig:viz}).
At the other times, the variations are within the 3$\sigma$ intervals.

The benefit of the cumulative count plot is that it is independent
of any binning. It stacks the information of neighbouring time bins.
Instead of modifying the data through background subtraction, the
data are fixed. A drawback is that time intervals cannot be investiaged
in isolation, as all time bins are correlated to the time bins before.
Furthermore, these visualisations are not rigorous statistical variability
tests, as the data will lie outside the $3\sigma$ regions occasionally,
given enough time bins.

\subsection{Methods for variability detection\label{subsec:Methods-for-variability}}

In this section, methods for detecting and quantifying variability
are contrasted. In eFEDS, the time series are sparsely sampled (see
§\ref{sec:Data}). Within a total length of only a few hours, each
source is typically continuously observed for a few minutes, about
$N\sim20$ times (less often at the edges of the field). In this setting,
we test several methods for their ability to detect variability, and
quantify how sensitive they are to different types of variability.

\subsubsection{Amplitude Maximum Deviation methods}

The simplest definition of variability is that two measured source
rates disagree with each other. This implies that the source has changed.

Assuming Gaussian error propagation, \citet{Boller2016} defined the
Amplitude Maximum Deviation (\emph{ampl\_max}) as the tension between
the most extreme points:
\begin{align}
\mathrm{AMPL\_MAX} & = & (\hat{R}_{S}(t_{\mathrm{max}})-\hat{\sigma}(R_{S})(t_{\mathrm{max}}))-\nonumber \\
 &  & (\hat{R}_{S}(t_{\mathrm{min}})+\hat{\sigma}(R_{S})(t_{\mathrm{min}}))\label{eq:MAD}
\end{align}
where $t_{\mathrm{min}}$ and $t_{\mathrm{max}}$ are the time bins
with the lowest and highest $\hat{R}$ source rate estimate. The ampl\_max
is the distance between the lower error bar of the maximum value to
the upper error bar of the minimum value. By comparing the span to
the error bars, the significance can be quantified in units of standard
deviations (i.e., as a z-score): 
\begin{equation}
\mathrm{AMPL\_SIG}=\frac{\mathrm{AMPL\_MAX}}{\sqrt{\hat{\sigma}(R_{S})(t_{\mathrm{max}})^{2}+\hat{\sigma}(R_{S})(t_{\mathrm{min}})^{2}}}\label{eq:MADsig}
\end{equation}

This method is conservative, as it considers the error bars twice.
A drawback of this method is the assumption that the errors are Gaussian.
With the asymmetric Poisson errors derived in section~\ref{sec:srcestimatesbayes},
one could define an analogous Bayesian AMPL\_MAX and AMPL\_SIG by
modifying eq.~\ref{eq:MADsig} to use the Bayesian quantile uncertainties
instead of $\hat{\sigma}(R)$. Such a modified method was considered
for the simulations performed in this paper. However, it yielded comparable
efficiency in detecting variability. This is probably because this
method is limited primarily by considering only the two extreme data
points, rather than by a refinement of the error bars.

The Amplitude Maximum Deviation quantifies both the size of the effect
(eq.~\ref{eq:MAD}) and its statistical significance (eq.~\ref{eq:MADsig}).
Because only the two most extreme data points are considered, it is
thus insensitive to the variations in the other values.

\subsubsection{Bayesian blocks}

The Bayesian blocks algorithm \citet{Scargle2013} identifies in a
sequence of measurement points where the rate changed. This adaptive
binning technique automatically segments a light curve into blocks
of constant rates separated by change points. The criterion to decide
the number and location of the change points is based on Bayesian
model comparison. For a certain class of likelihood functions, \citet{Scargle2013}
derived analytic recursive formulas which quickly construct the globally
optimal segmentation. Bayesian blocks can be applied to photon counts
of binned light curves and even to individual photon count arrival
times, thus not requiring a pre-defined binning. However, as Figure~\ref{fig:rawdata}
illustrates, the eROSITA photon counts are highly variable as the
source runs through the field-of-view in the survey scan, simply because
of angle-dependent instrument sensitivity. For astrophysical inference,
we are interested in source variability, rather than observation-induced
variability. The Bayesian blocks algorithm could be extended with
a new likelihood to incorporate this information. However, a further
difficulty is that the background is not negligible for most sources.
Some of its components are variable over time, especially those passing
through the mirrors and those sensitive to spacecraft orientation
relative to the sun. Others, such as the particle background, are
persistent, and become dominant at large off-axis angles. An extension
of Bayesian blocks to analyse source and background region light curve
simultaneously would be desirable, building on the foundations outlined
above. However, this is beyond the scope of this work. Therefore,
we resort to the classic source rate and uncertainty estimators $\hat{R}(t)$
and $\hat{\sigma}(\mathrm{R})(t)$, which are corrected for the fractional
exposure, and use the Gaussian Bayesian blocks implementation from
astropy \citep{AstropyCollaboration2013,AstropyCollaboration2018}.
This however requires pre-binned light curves.

In our application to binned light curves, all borders between time
bins with observations are candidates for change points. Bayesian
blocks begins with the hypothesis that the count rate is constant.
For each candidate change point, it tries the hypothesis that the
count rate is constant to some value before the change point, and
constant to some value after the change point. The two hypothesis
probabilities are compared using Bayesian model comparison. If the
model comparison favours the split, each segment is analysed with
the same procedure recursively. Finally, Bayesian blocks returns a
segmented light curve, and estimates for the count rate in each segment
with its uncertainties.

The Bayesian model comparison requires a prior on the expected number
of change points $n_{\mathrm{cp}}$. We adopt the prior favoured by
simulations of \citet{Scargle2013}, $P(n_{\mathrm{cp}})=4-73.53p_{0}n_{\mathrm{cp}}^{-0.478}$
with the desired false positive rate set to $p_{0}=0.003$ (corresponding
to 3$\sigma$). Variability is significantly ``detected'' by the
Bayesian blocks algorithm when it identified at least one change point.
We refer to $n_{\mathrm{cp}}$ as NBBLOCKS.

\subsubsection{Fractional and Excess variance}

A Poisson process is expected to induce stochasticity into the measurement.
Excess variance methods \citep{Edelson1990,Nandra1997,Edelson2002,Vaughan2003,Ponti2014}
quantify whether the observed stochasticity shows additional variance,
i.e., is over-dispersed.

Across bins, the mean net source count rate $\bar{R}_{S}$ is: 
\begin{equation}
\bar{R}_{S}={\frac{1}{N}\sum_{i}^{N}{\hat{R}_{S}(t_{i})}}\label{eq:obsmean}
\end{equation}
The observed variance of the net source count rates $\hat{R}_{S}$
(one in each time bin) is:
\begin{equation}
\sigma_{\mathrm{obs}}^{2}=\frac{1}{N-1}\sum_{i}^{N}{\left(\hat{R}_{S}(t_{i})-\bar{R}_{S}\right)^{2}}\label{eq:obsvar}
\end{equation}
The Poisson noise expectation is computed with the mean square error
computed from the error bars: 
\begin{equation}
\overline{\sigma_{\mathrm{err}}^{2}}=\frac{1}{N}\sum_{i}^{N}{\left(\hat{\sigma}(R_{S})(t_{i})\right)^{2}}\label{eq:poissonvar}
\end{equation}
Subtracting off this expectation, we obtain the excess variance:
\begin{equation}
\sigma_{\mathrm{XS}}^{2}=\sigma_{\mathrm{obs}}^{2}-\overline{\sigma_{\mathrm{err}}^{2}}\label{eq:exvar}
\end{equation}
Normalising to the mean count rate, gives the normalised excess variance
(NEV): 
\begin{equation}
\mathrm{NEV}=\frac{\sigma_{\mathrm{XS}}^{2}}{\bar{R}_{S}^{2}}\label{eq:nev}
\end{equation}
The variable fraction of the signal, $F_{\mathrm{var}}$, also known
as the fractional root-mean-square (RMS) amplitude, is then defined
as:
\begin{equation}
F_{\mathrm{var}}=\sqrt{\mathrm{NEV}}\label{eq:fvar}
\end{equation}

The excess variance $\sigma_{\mathrm{XS}}^{2}$ quantifies the over-dispersion,
without making assumptions about the process causing the variability.
Values of $\sigma_{\mathrm{XS}}^{2}$ can however also become negative
by chance, or when the measurement uncertainties are over-estimated.
To avoid this problem (which affects $F_{\mathrm{var}}$), we force
NEV to not go below a small positive value (0.001).

Quantifying the significance of the excess variance is more difficult
\citep{Nandra1997}. \citep{Vaughan2003} used simulations to find
the empirical formulas (valid for $N$ from 2 to 2000 and $F_{\mathrm{var}}$
from 0 to $40\%$) for the uncertainty in the $\mathrm{NEV}$ and
$F_{\mathrm{var}}$ estimators:
\begin{equation}
\sigma(\mathrm{NEV})=\sqrt{\frac{2}{N}\left(\frac{\overline{\sigma_{\mathrm{err}}^{2}}}{\bar{R}_{S}^{2}}\right)^{2}+\frac{\overline{\sigma_{\mathrm{err}}^{2}}}{N}\times\left(\frac{2\times F_{\mathrm{var}}}{\bar{R}_{S}}\right)^{2}}\label{eq:signev}
\end{equation}

\begin{equation}
\sigma(F_{\mathrm{var}})=\frac{\sigma(\mathrm{NEV})}{2\times F_{\mathrm{var}}}\label{eq:sigfvar-1}
\end{equation}
The significance of the excess variance can then be defined as $\mathrm{FVAR\_SIG}=F_{\mathrm{var}}/\sigma(F_{\mathrm{var}})$
and $\mathrm{NEV\_SIG}=\mathrm{NEV}/\sigma(\mathrm{NEV})$.

\subsubsection{Bayesian excess variance (bexvar)}

\label{subsec:Bayesian-excess-variance}
\begin{figure}
\centering{}\includegraphics[width=0.8\columnwidth]{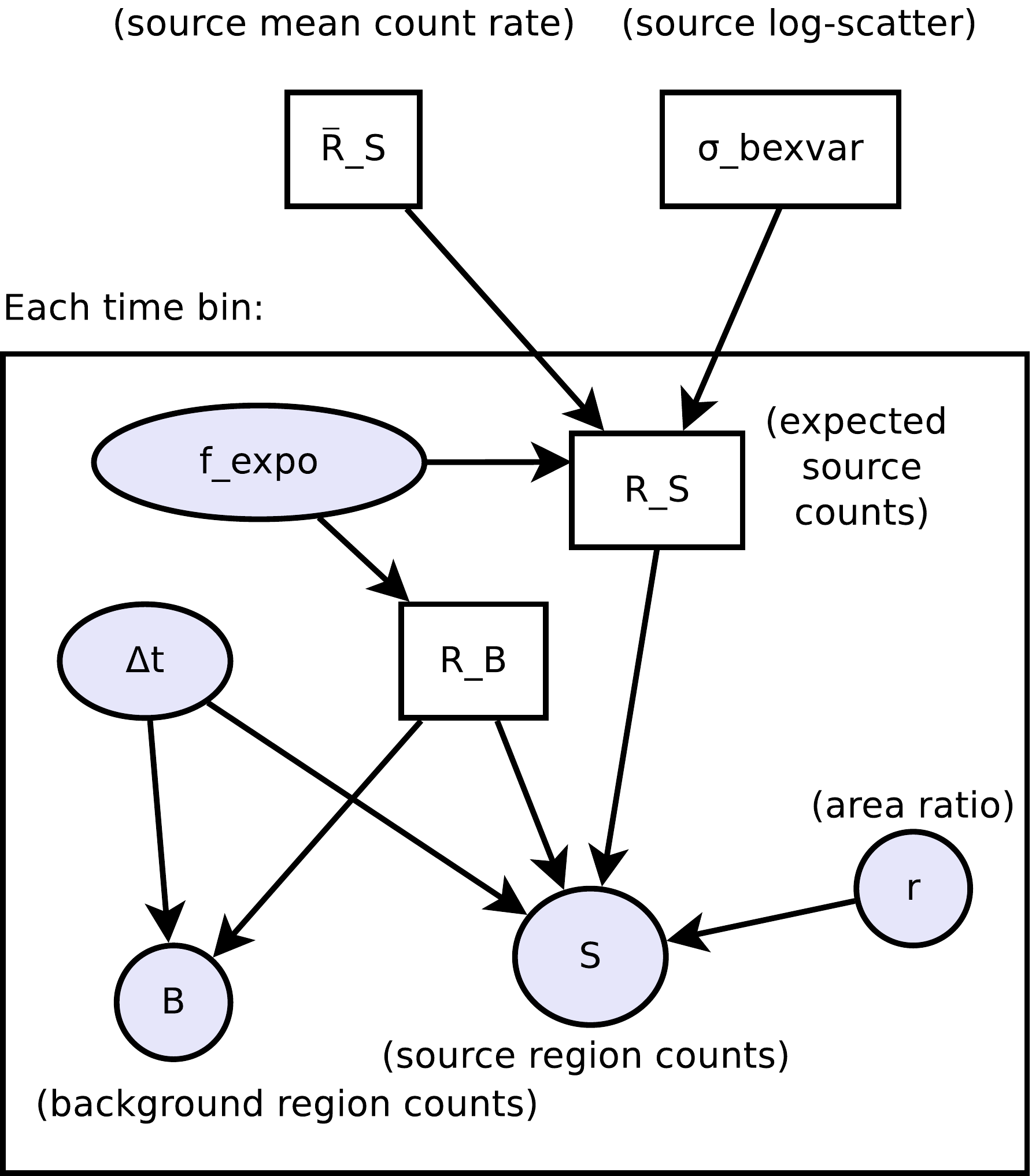}\caption{Graphical model of the Bayesian excess variance method. Shaded circles
indicate known values, related to the experiment setup or observed
data. Rectangles indicate unknown parameters, including the unknown
source count rate and the count rate in each time bin. An arrow from
A to B indicates that the generation of B was influenced by A.}
\label{fig:hbmviz}
\end{figure}

The excess variance computation above assumes symmetric, Gaussian
error bars. This limitation can be relaxed by modelling the entire
data generating process. Towards this, we assume that at any time
bin $i$, the rate $R_{S}(t_{i})$ is distributed according to a log-normal
distribution with unknown parameters: 
\begin{equation}
\log R_{S}(t_{i})\sim\mathrm{Normal}\left(\log\bar{R}_{S},\sigma_{\mathrm{bexvar}}\right)\label{eq:hbm}
\end{equation}
In this formulation, we need to estimate the mean logarithmic net
source count rate ($\log\bar{R}_{S}$), the intrinsic scatter $\sigma_{\mathrm{bexvar}}$
as well as the rates at each time bin $R_{S}(t_{i})$, giving $N+2$
parameters. Equation~\ref{eq:hbm} defines a prior for each bin's
source count rate. This is a hierarchical Bayesian model (HBM), combined
with the equations \ref{eq:bkgpoisson} and \ref{eq:srcpoisson} which
define the probabilities in each time bin. Figure~\ref{fig:hbmviz}
illustrates the relation between all quantities as a graphical model.

Priors for $\bar{R}_{S}$ and $\sigma_{\mathrm{bexvar}}$ also need
to be chosen. Here, we simply use uninformative, wide flat priors:
\begin{flalign}
\log\bar{R}_{S}\sim\mathrm{Uniform}\left(-5,5\right)\label{eq:hbmpriorR}\\
\log\sigma_{\mathrm{bexvar}}\sim\mathrm{Uniform}\left(-2,2\right)\label{eq:hbmpriorS}
\end{flalign}
The mean count rate $\bar{R}_{S}$, has a straight-forward interpretation.
In its posterior distribution, the Poisson uncertainty is directly
incorporated.

Variability is quantified with $\sigma_{\mathrm{bexvar}}$, which
gives the intrinsic variance. This log-scatter on the log-count rate
is a different quantity than the excess variance on the (linear) count
rate, $\sigma_{XS}$. Because the variability is defined as a log-normal,
this corresponds to the log-amplitude of a multiplicative process.
The motivation for this is primarily of practical. Variable objects
can be identified when the posterior distribution of $\sigma_{\mathrm{bexvar}}$
excludes low values. Here, we define SCATT\_LO as the lower 10\% quantile
of the posterior, and use it as a variability indicator.

\begin{figure}
\includegraphics[width=1\columnwidth]{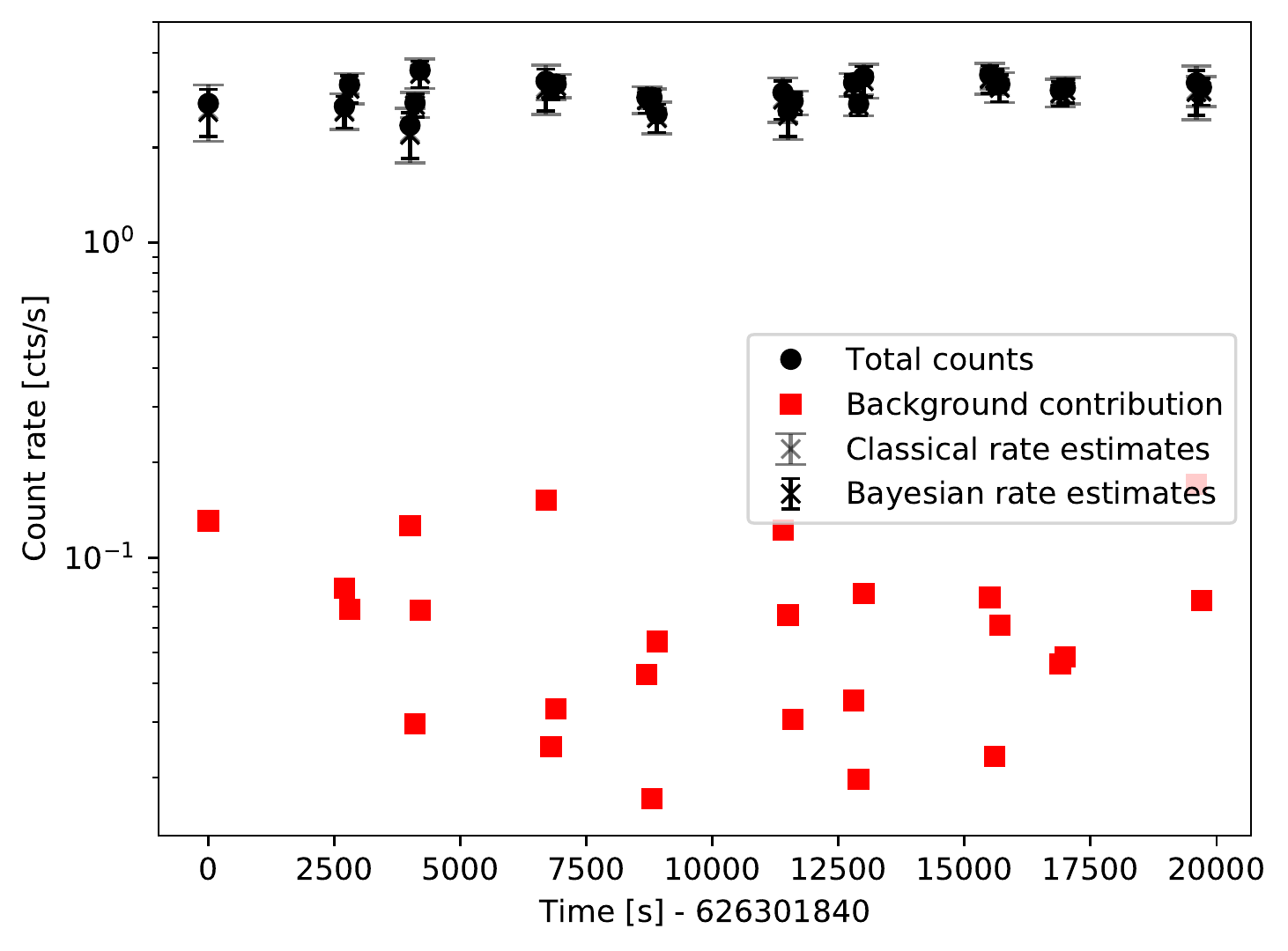}\\
 \includegraphics[width=1\columnwidth]{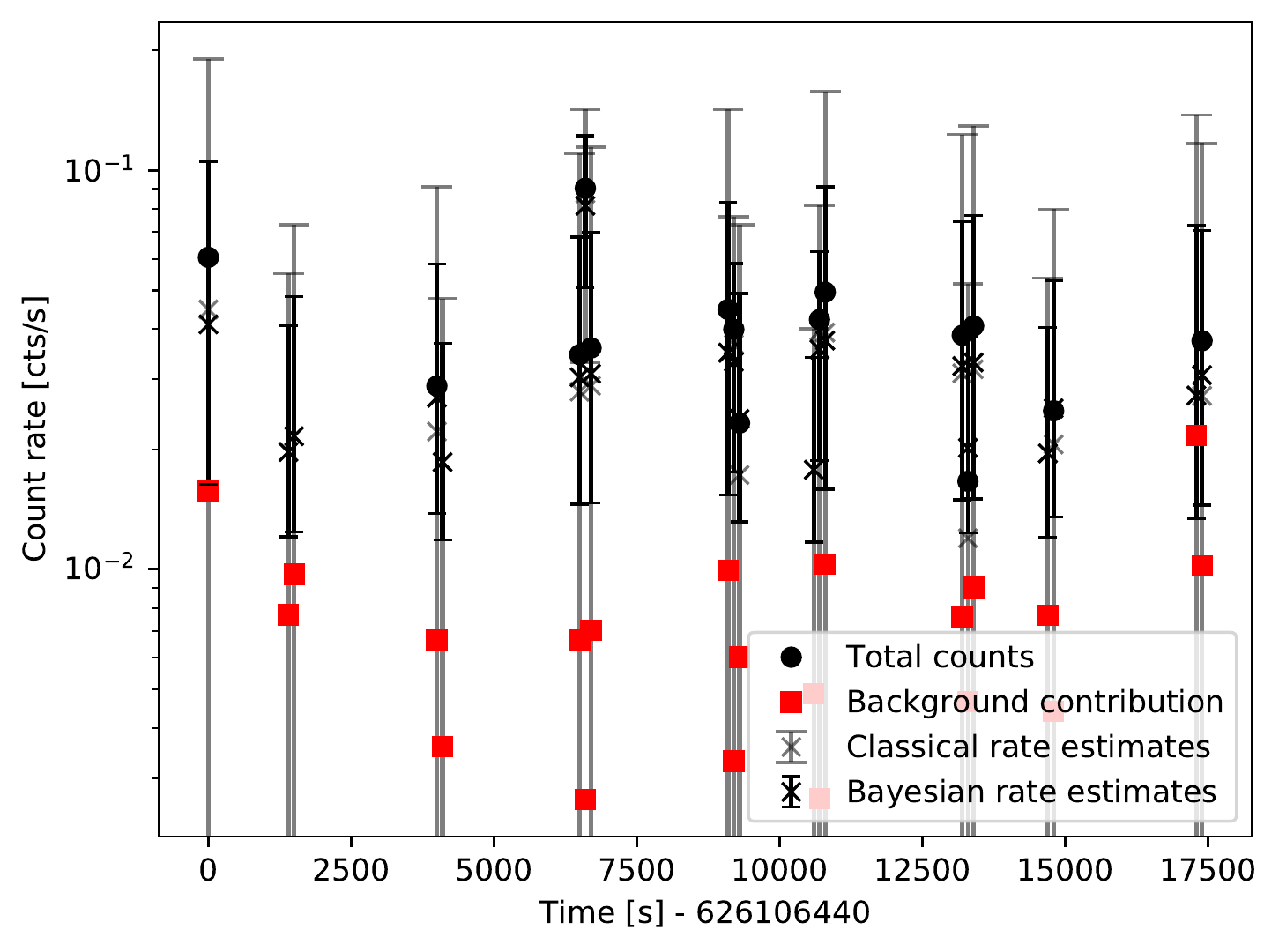}
\caption{Simulated light curves of \textit{constant} sources with Poisson noise.
Black circles show the total counts without background subtraction,
red points show the expected background count rates in the source
region (eq.~\ref{eq:bkgrate}). The \textit{top panel} shows a high-count
rate source with a constant 3 cts/s, where the source counts are always
above the expected background counts. The \textit{bottom panel} shows
a low-count rate source with a constant 0.03 cts/s, where source region
counts and background counts are comparable and show substantial Poisson
scatter. Gray error bars show Classical net source count rate estimates
(eq.~\ref{eq:srcrate} and \ref{eq:srcrateerr}). Black error bars
show Bayesian net source count rate posterior distributions (eq.~\ref{eq:pobjnumericalinv}).}
\label{fig:simlcexamplesconst}
\end{figure}

\begin{figure*}
\centering \includegraphics[width=1\columnwidth]{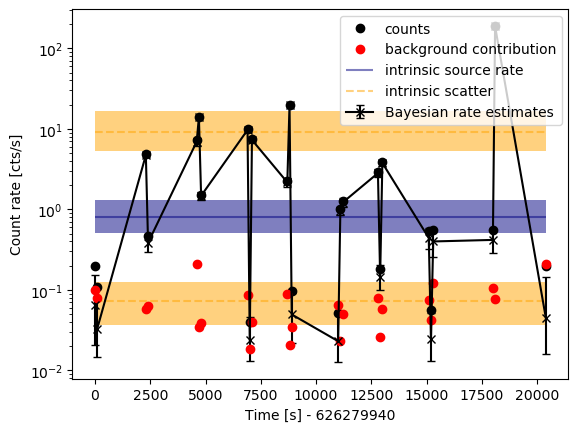}
\includegraphics[width=1\columnwidth]{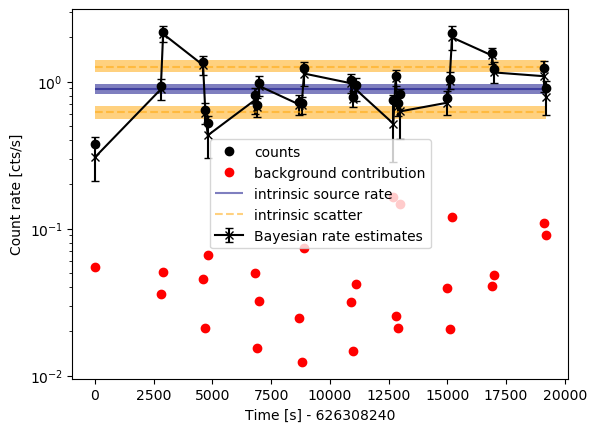}

\includegraphics[width=1\columnwidth]{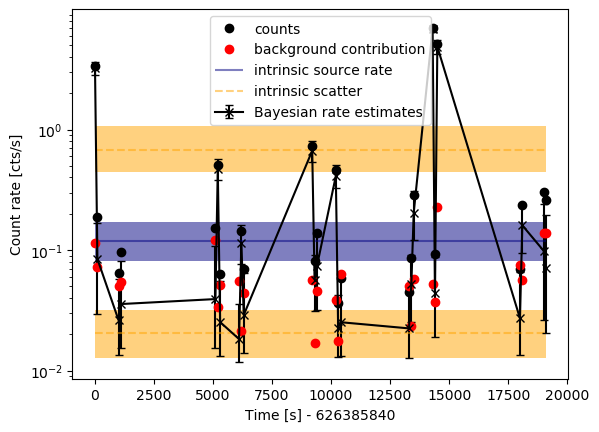}
\includegraphics[width=1\columnwidth]{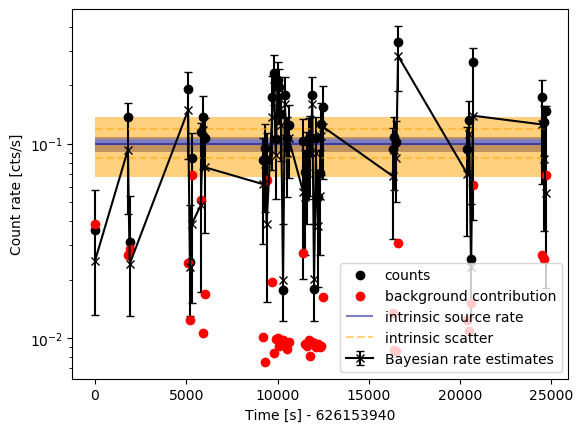}
\caption{As in Fig.~\ref{fig:simlcexamplesconst}, but for \textit{gaussvar}
sources. The intrinsic count rate is randomly varied following a log-normal
distribution around a base-line count rate. Left panels show cases
with $\sigma=1\mathrm{dex}$ variations, clearly visible in the scatter
of the total counts (black points) in both bright (top panel) and
faint (bottom panel) sources. Right panels show lower variations ($\sigma=0.1\mathrm{dex}$).
In the bright case (top panel) the scatter of the black points is
substantially larger than the error bars, while in the faint case
(bottom) error bars overlap. In the top panel the background contribution
(red points) is well below the total counts (black), while in the
bottom panel, they are comparable. The blue solid line and band shows
the posterior median and $1\sigma$ uncertainty of the intrinsic source
count rate $\hat{R}$, computed using the bexvar method. The intrinsic
scatter around the mean, $\sigma_{\mathrm{bexvar}}$, is shown in
orange dashed lines, indicating the upper and lower $1\sigma$ of
the estimated log-Gaussian. The uncertainties on $\sigma_{\mathrm{bexvar}}$
are shown in orange bands. In all but the bottom right panel, the
orange band is clearly separated from the blue band (indicating significant
intrinsic variability).}
\label{fig:simlcexamplesgauss}
\end{figure*}

\begin{figure*}
\centering \includegraphics[width=1\columnwidth]{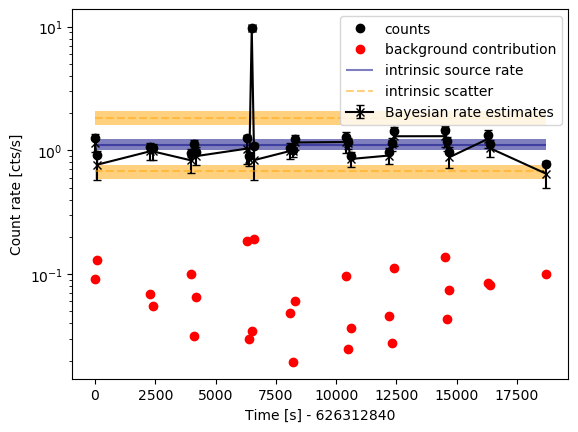}
\includegraphics[width=1\columnwidth]{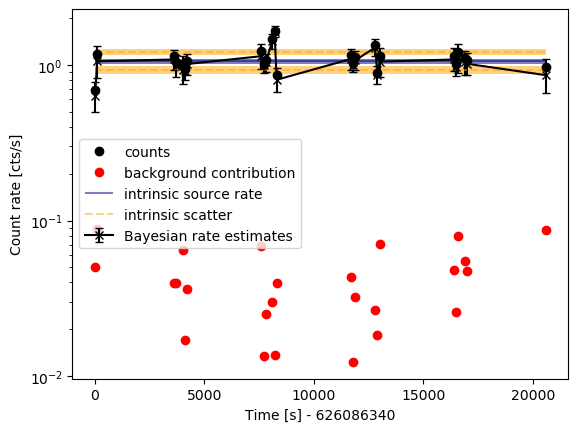}

\includegraphics[width=1\columnwidth]{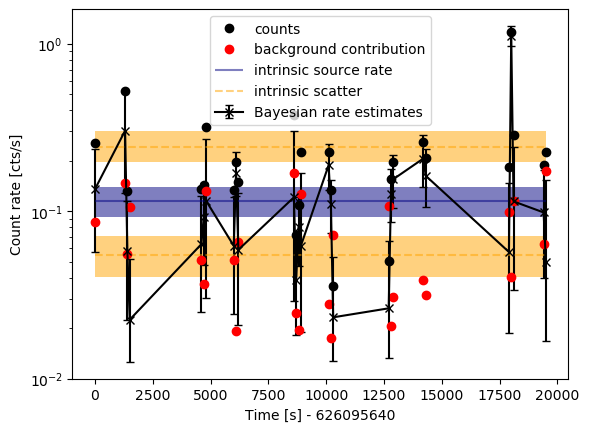}
\includegraphics[width=1\columnwidth]{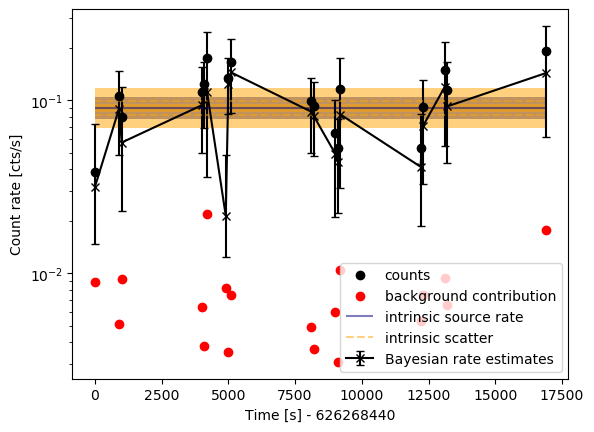}
\caption{As in Fig.~\ref{fig:simlcexamplesgauss}, but for \textit{flare}
sources. One time bin has its count rate increased by a factor of
$k$. Top (bottom) panels show bright (faint) source base-line count
rates. Top panels show a flare of $k=10$, bottom panels a $k=2$
flare.}
\label{fig:simlcexamplesflare}
\end{figure*}

How can the above formulas be solved to actually produce probability
distributions on, say, $\sigma_{\mathrm{bexvar}}$? The first step
is the posterior probability computation. If we assume a source count
rate $R_{S}(t_{i})$ at each bin, eq.~\ref{eq:srcpoisson} indicates
how to compute the Poisson probability to detect the source region
counts $S$. The background count rate $R_{B}(t_{i})$ also needs
to be chosen and the Poisson probability to detect the background
region counts $B$ can be computed. If we further assume a value $\bar{R}_{S}$
and $\sigma_{\mathrm{bexvar}}$, eq.~\ref{eq:hbm} computes the probability
of the $N$ chosen $R_{S}(t)$ values. Finally, eq.~\ref{eq:hbmpriorR}
and \ref{eq:hbmpriorS} specify the prior probability for $\bar{R}_{S}$
and $\sigma_{\mathrm{bexvar}}$, i.e., $\pi(\bar{R}_{S})$ and $\pi(\sigma_{\mathrm{bexvar}})$,
respectively. To summarise, given the assumed $(2\times N+2)$-dimensional
parameter vector we compute $2N+3$ probabilities for light curve
data $D=(S_{1},B_{1},...,S_{N},B_{N})$:
\[
\theta=(\bar{R}_{S},\sigma_{\mathrm{bexvar}},R_{S}(t_{1}),...,R_{S}(t_{N}),R_{B}(t_{1}),...,R_{B}(t_{N}))
\]
As we require all probabilities to hold simultaneously, we multiply
them, obtaining the posterior probability function: 
\begin{equation}
P(\theta|D)=\pi(\bar{R}_{S})\times\pi(\sigma_{\mathrm{bexvar}})\times\prod_{i=1}^{N}P(R_{S}(t_{i}),R_{B}(t_{i})|S_{i},B_{i})\label{eq:hbmpost}
\end{equation}
where the per-bin posterior probability terms:

\begin{multline}
P(R_{S}(t_{i}),R_{B}(t_{i})|S_{i},B_{i})=\\
P(R_{S}(t_{i})|\bar{R}_{S},\sigma_{\mathrm{bexvar}})\times\pi(R_{B}(t_{i}))=
P(S_{i},B_{i}|R_{S}(t_{i}),R_{B}(t_{i}))\label{eq:hbmpostlower-1}
\end{multline}
and the per-bin likelihoods being the product of equations \ref{eq:bkgpoisson}
and \ref{eq:srcpoisson}:
\begin{flalign}
P(S,B|R_{S},R_{B})= & P(B|R_{B})\times P(S|R_{S},R_{B})\label{eq:hbmpostlower}
\end{flalign}
We are now equipped with a posterior over a $(2\times N+2)$-dimensional
parameter space. To compute probability distributions for a parameter
of interest, say, $\sigma_{\mathrm{bexvar}}$, all other parameters
need to be marginalised out:
\begin{equation}
P(\sigma_{\mathrm{bexvar}}|D)=\int\int\int P(\theta|D)\times d\bar{R}_{S}dR(t_{i})dR_{B}(t_{i})
\end{equation}
The exploration of the posterior probability distribution on $\sigma_{\mathrm{bexvar}}$
can be achieved with Markov Chain Monte Carlo algorithms. These repeatedly
propose values $\theta$, and sample the values of $\sigma_{\mathrm{bexvar}}$
proportional to their posterior probability. However, in practice,
the convergence of this computation is slow and not always stable,
even with state-of-the-art methods.

Substantial improvements are possible for rapid computation. Firstly,
section~\ref{sec:srcestimatesbayes} already derived the marginalised
likelihood for the per-bin source rates $P(S,B|R_{S})$. These can
be represented as an array $P_{i,j}$ giving the probability for time
bin $i$ over a grid of source rates $R_{j}$. Using the normal distribution
defined by $\bar{R}_{S}$ and $\sigma_{\mathrm{bexvar}})$, we can
compute at each grid $R(t)$ value its probability, and marginalise
over $j$ over the grid for each time bin. Thus, we approximate eq.~\ref{eq:hbmpost}
with: 
\begin{multline}
P(D|\bar{R}_{S},\sigma_{\mathrm{bexvar}})\propto \\ \prod_{i=1}^{N}{\sum_{j=1}^{M}P_{i,j}\times\mathrm{Normal}(\log R_{j}|\log\bar{R}_{S},\sigma_{\mathrm{bexvar}}))}\label{eq:hbmpostnumerical}
\end{multline}
We are now left with only a two-dimensional probability distribution.
We employ the nested sampling Monte Carlo algorithm MLFriends \citep{buchner2016statistical,Buchner2019c}
implemented in the UltraNest Python package\footnote{\url{https://johannesbuchner.github.io/UltraNest/}}
\citep{Buchner2021} to obtain the probability distribution $P(\sigma_{\mathrm{bexvar}}|D)$
using the likelihood (eq.~\ref{eq:hbmpostnumerical}) and priors
(eq.~\ref{eq:hbmpriorR} and \ref{eq:hbmpriorS}).

The probability distribution $P(\sigma_{\mathrm{bexvar}}|D)$ quantifies
the variability amplitude supported by the data. To illustrate the
typical behaviour of this probability distribution, once a few data
points are added, the highest values of $\sigma_{\mathrm{bexvar}}$
are excluded and receive a low probability. When inconsistent data
points (excess variance) is present, also the lowest values of $\sigma_{\mathrm{bexvar}}$
receive a low probability, concentrating the probability distribution
near the true value. Therefore, a conservative indicator of the magnitude
of the excess variance is the lower 10\% quantile of the distribution,
which we term SCATT\_LO. We adopt this single summary statistic for
comparison with the other methods. Similar to the other methods (NEV,
Bayesian blocks), the significance quantification needs to be obtained
with simulations.

\subsection{Method comparison\label{subsec:Method-comparison}}

To summarise, we consider four methods (and their estimators): 
\begin{itemize}
\item Amplitude Maximum Deviation (AMPL\_MAX) 
\item Normalised Excess Variance (NEV)
\item Bayesian Blocks (NBBLOCKS)
\item Bayesian Excess Variance (SCATT\_LO)
\end{itemize}
All of these methods rely on binned light curves. They are therefore
sensitive to the chosen number of bins, which modulates how much information
is contained in each bin. All methods neglect time information and
are oblivious to the order of measurements. The exception is Bayesian
blocks. All methods disregard gaps.

To quantify the significance of a detection, the first three methods
have already a significance indicator (AMPL\_SIG, NEV\_SIG, NBBLOCKS).
However, these are derived under specific assumptions or simulation
settings: The existing simulations of \citet{Vaughan2003} and \citep{Scargle2013}
did not consider the scenario with variable sensitivity and non-negligible
backgrounds. The AMPL\_SIG does not account for the number of data
points, which increase the chance of getting a large AMPL\_MAX by
chance. The Bayesian blocks, NEV and AMPL\_SIG methods adopt imperfect
Gaussian approximations. Because of these limitations, simulations
are necessary to detect variable objects with desired reliability
characteristics. For these reasons, we verify that the significance
indicators correspond to the desired p-values, for example, in constant
sources NEV\_SIG should exceed $3\sigma$ only in $0.1\%$ of cases
by chance. We prefer to verify NEV\_SIG and AMPL\_SIG rather than
NEV and AMPL\_MAX, as the existing formulae already largely correct
for trends with size of the uncertainties and the number of data points.
In Bayesian blocks, unjustified change points should also rarely introduced
by chance.

\subsection{Simulation setup\label{subsec:Simulation-setup}}

To calibrate the significance threshold at which an object is classified
as variable, we use extensive simulations. Four datasets are generated.
Each data set is created based on the 27910 eFEDS light curves, taking
their time sampling ($\Delta T$) and $f_{\mathrm{expo}}$ values
as is. This generates a data set under identical conditions. We do
not vary the vignetting and other corrections, i.e., assume that the
instrument model is correct. Background counts are sampled using equation~\ref{eq:bkgpoisson}
with Poisson random numbers assuming the measured, time-average $\hat{R_{B}}$
as a constant across all time bins for that source. Source counts
are sampled using equation~\ref{eq:srcpoisson} with Poisson random
numbers, using the sum of the scaled background rate and the desired
source rate $R(t)$ at that time step. For reasonable ranges of $R(t)$,
recall that the typical number of counts in a $100\,s$ bin is below
10 for most sources and time bins, but can reach up to a few hundreds
(see Fig.~\ref{fig:fieldstats-counts} and \citealp{2021arXiv210614523B}).
We therefore consider count rate ranges between $\mu=0.03$ cts/s
and 3 cts/s.

For the source rate, four scenarios are considered:
\begin{enumerate}
\item \textit{constant}: The count rate are constant: $R(t)=\mu$. The sample
is divided into five equally sized groups. Each group is assigned
a different count rate (0.03, 0.1, 0.3, 1 and 3 cts/s). Figure~\ref{fig:simlcexamplesconst}
presents two examples of \textit{constant} light curves, showing the
lowest and highest count rates considered. The Poisson scatter is
strongly noticeable.
\item \textit{gaussvar}: For each time bin, a count rate is drawn independently
from a log-normal distribution\textbf{ }with mean $\log\mu$ and variance
$\sigma$: $\log R(t)\sim\mathrm{Normal}(\log\mu,\,\sigma$). The
sample is divided into five equally sized groups. Each group is assigned
a different mean count rate ($\mu=$0.03, 0.1, 0.3, 1 and 3 cts/s).
The groups are further subdivided into five subgroups. Each subgroup
is assigned a different variance ($\sigma=0.03,$0.1, 0.3, 0.5, 1.0).
This represents the behaviour of long-term revisits of AGN \citep[e.g.,][]{Maughan2019}.
Figure~\ref{fig:simlcexamplesgauss} presents four examples of \textit{gaussvar}
light curves, varying the intrinsic count rates (left vs. right panels)
and the strength of the intrinsic scatter (top vs. bottom panels).
These examples also illustrate the inference of bexvar, estimating
the mean, intrinsic count rate (blue line), its uncertainty (blue
band), the count rate log variation (orange lines) and its uncertainty
(orange bands). 
\item \textit{flare}: Same as \textit{constant}, but in one randomly selected
time bin the count rate is increased by a factor $k$. The groups
are subdivided into subgroups. Each subgroup is assigned a different
factor ($k=30,$10, 5, 2, 1.5, 1.3). The subgroups with $k=$30 and
5 are half as large as the other subgroups. Figure~\ref{fig:simlcexamplesflare}
presents four examples of \textit{flare} light curves, varying the
intrinsic count rates (left vs. right panels) and the flare strength
$k$ (top vs. bottom panels). Weak flares become difficult to notice
in the presence of Poisson noise.
\item \emph{redvar}: To complement the white noise process \textit{\emph{in}}\textit{
gaussvar}, a correlated random walk (red noise) is also tested. Specifically,
we adopt a first-order Ornstein--Uhlenbeck process $x_{i+1}=\phi\times x_{i}+\mathrm{\Delta t\times Normal(0,1)}$.
This generates on short time-scales a powerlaw power spectrum with
index -2, not atypical of AGN \citep[e.g.,][]{Simm2016}. To always
be near this regime, we choose a long dampening time-scale $T=100,000\,\mathrm{s}$,
so that $\phi=\exp\left(-\Delta t/T\right)$ is close to 1. The long-term
variance of this random walk is $\sigma^{2}=\Delta t^{2}/(1-\phi^{2})$.
Following \citet{Vaughan2003}, time bins are super-sampled forty-fold
to avoid red noise leaks, and then summed. Finally, the random walk
is normalised and mixed as a variable fraction $f_{\mathrm{var}}$
with a constant to obtain the source count rate as $R(t_{i})=\mu\times(1+f_{\mathrm{var}}\times x_{i}/\sigma)$.
The mean $\mu$ is varied in groups as in \textit{gaussvar}, and five
equally sized subgroups set $f_{\mathrm{var}}$ to $0.03,$0.1, 0.2,
0.3 and 0.5, motivated by the range observed in X-ray binaries \citep{Heil2015}. 
\end{enumerate}
Simulating other types of variability, such as exponential or linear
declines, sinusoidal variations are outside the scope of this work.
However, because almost all methods adopted here ignore the order
of measurements, they are covered to some degree by the \textit{gaussvar}
setup.

To achieve accurate quantification of the methods, many simulations
are needed. In total, 27908 \textit{constant} simulations\textit{\emph{
(all eFEDS sources are used as templates)}}\textit{, }\textit{\emph{14003
}}\textit{gaussvar}\textit{\emph{ simulations (eFEDS sources with
even IDs are templates), 13905 }}\textit{flare}\textit{\emph{ simulations
(eFEDS sources with odd IDs are templates) and 14003 }}\textit{redvar}\textit{\emph{
simulations (same as }}\textit{gaussvar}\textit{\emph{) were generated
for the soft band. A similar number of simulations was performed for
the hard band, except 28 light curves have no valid time bins and
were discarded. We primarily focus on the soft band simulations.}}

\begin{figure}
\centering \includegraphics[width=1\columnwidth]{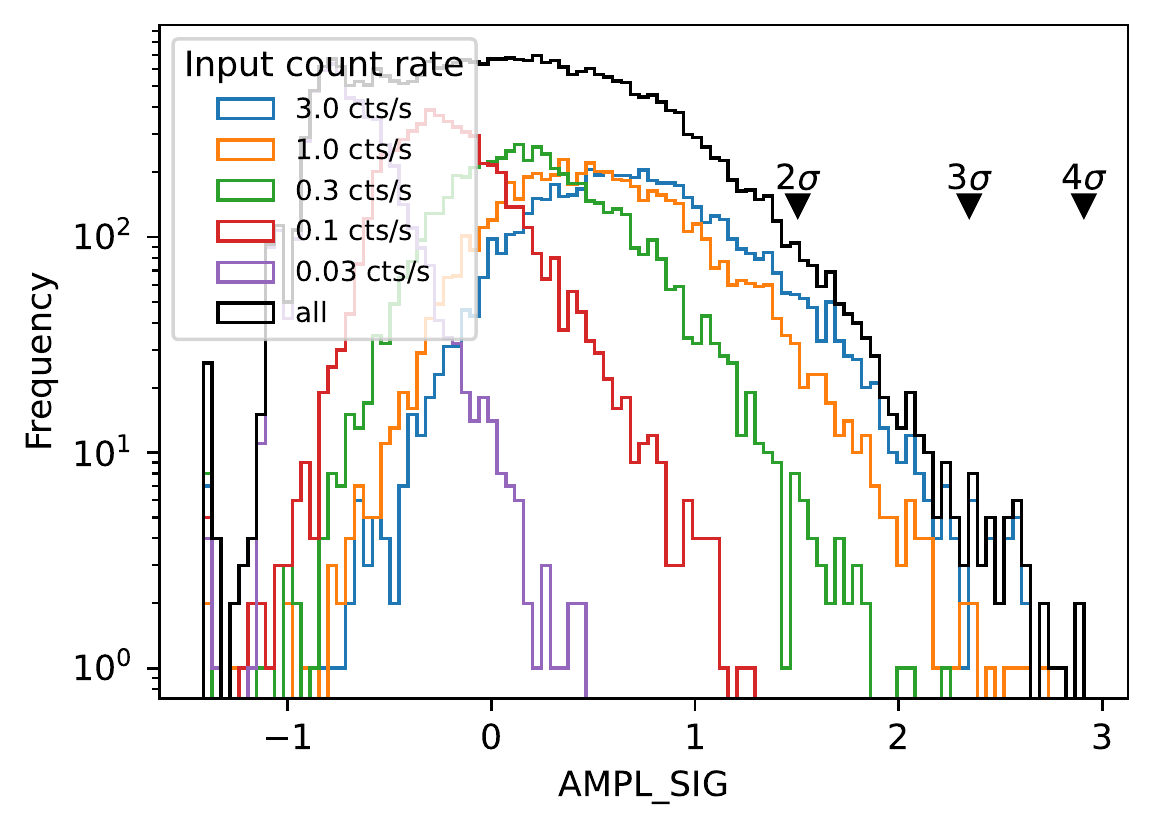} 

\includegraphics[width=1\columnwidth]{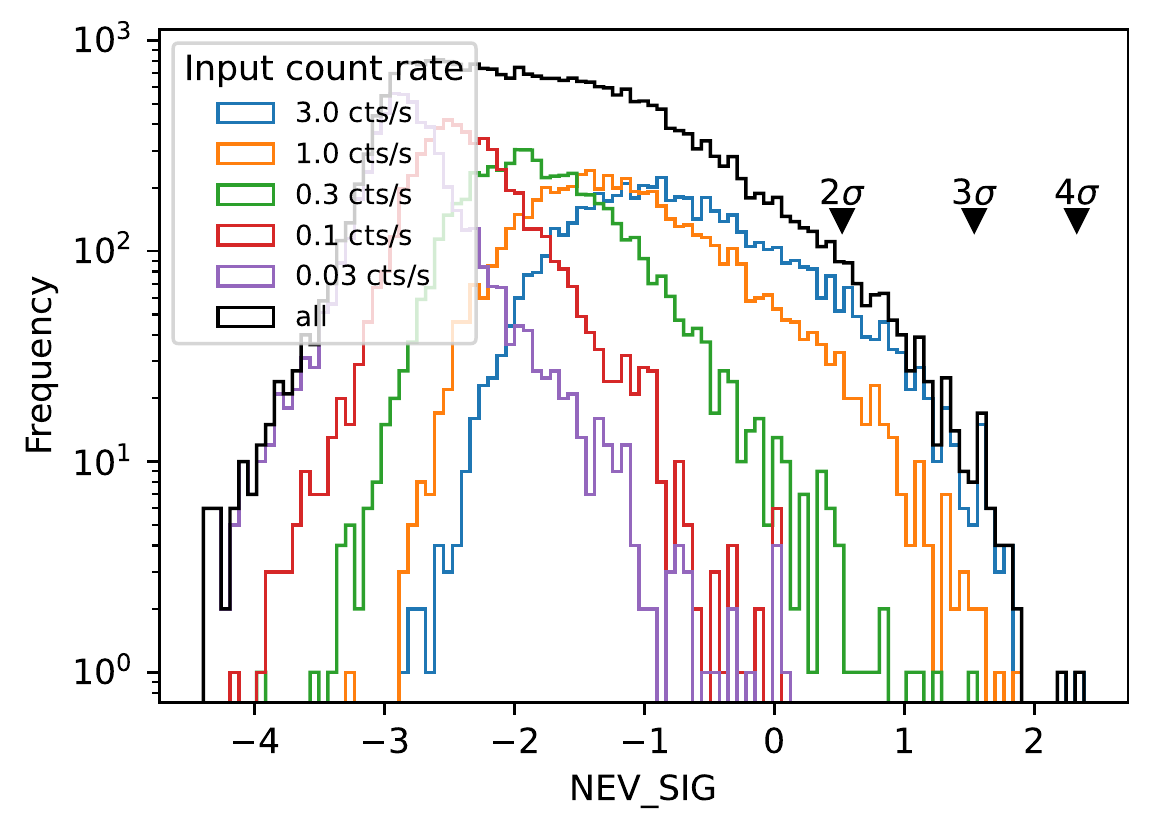}

\caption{Estimator distribution for simulated \textit{constant} light curves.
\textit{Top panel}: amplitude maximum deviation significance (eq.~\ref{eq:MADsig}).
\textit{Bottom panel}: Significance of the excess variance and variability
fraction (eq.~\ref{eq:signev}). Each coloured histogram represents
a set of simulations with the indicated constant input count rate.
For the full data set (black histogram), black downwards triangles
point to the $2\sigma$, $3\sigma$ and $4\sigma$ equivalent quantiles
of the distribution.}
\label{fig:simdists1}
\end{figure}

\begin{figure}
\centering \includegraphics[width=1\columnwidth]{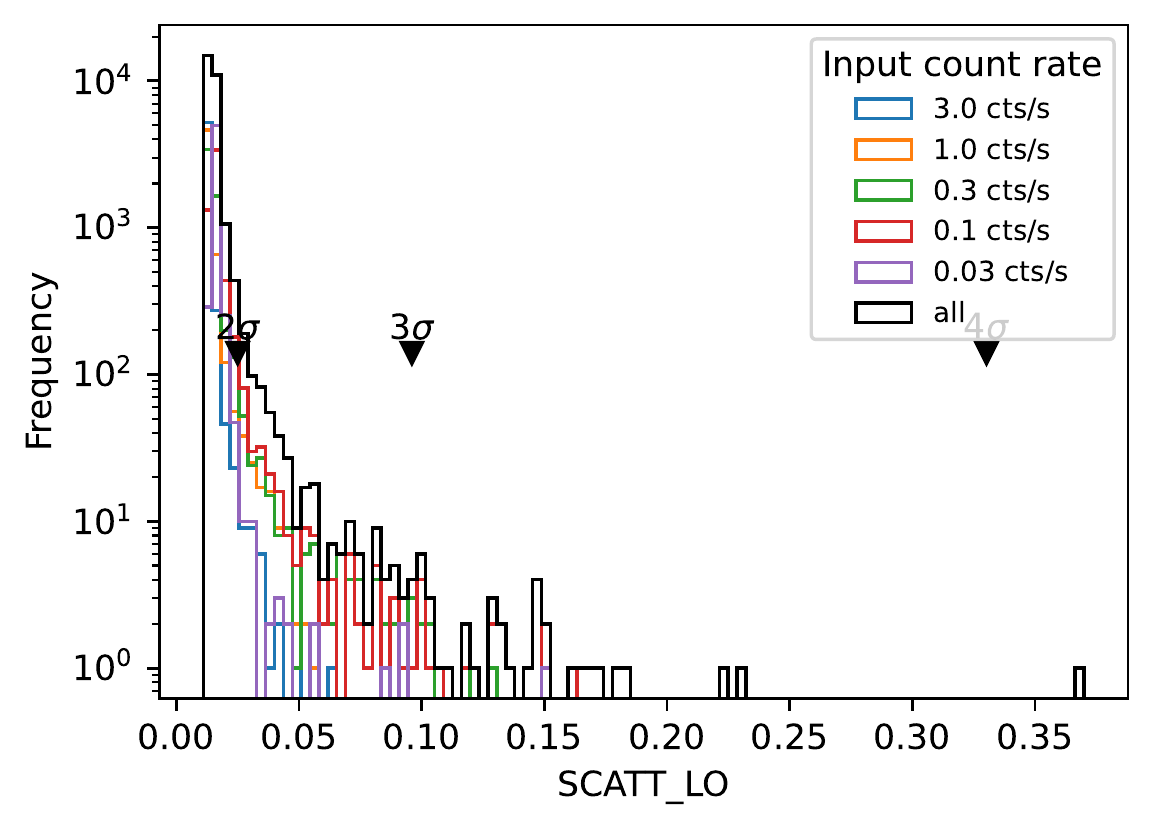} 

\includegraphics[width=1\columnwidth]{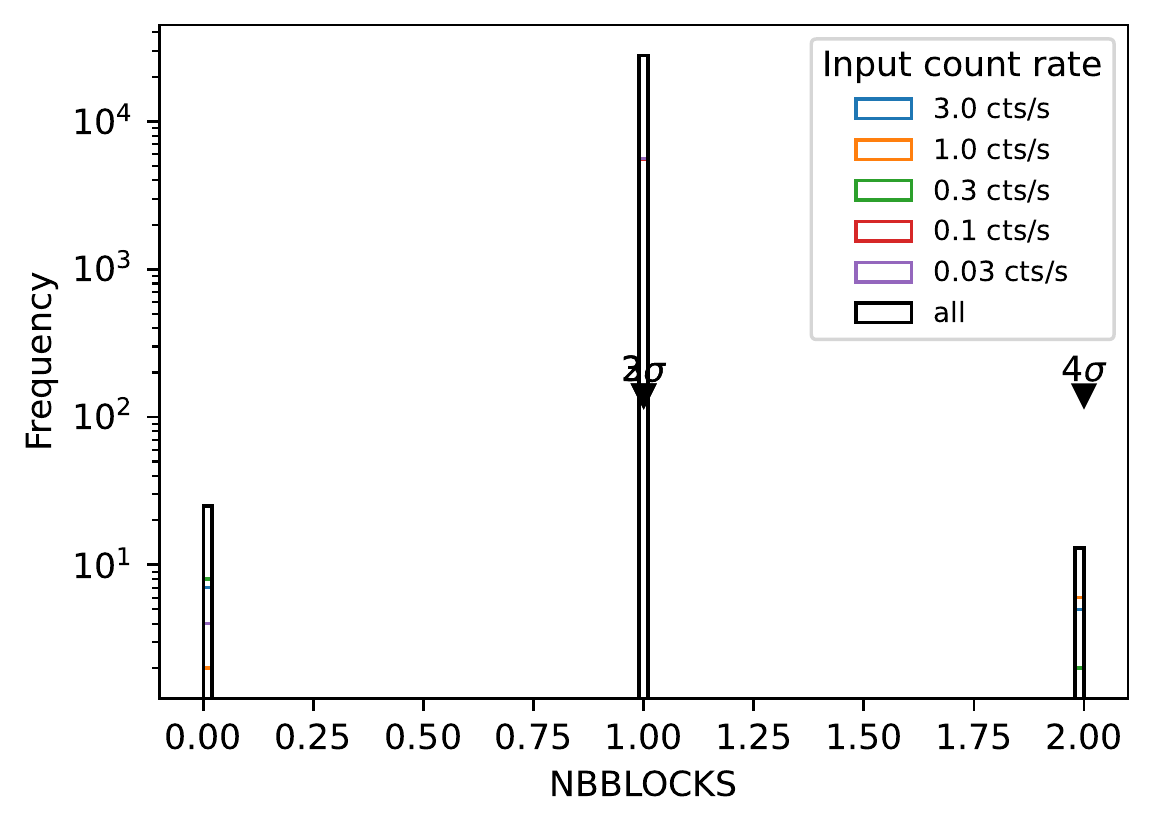}

\caption{As in Fig.~\ref{fig:simdists1}, but for the bexvar SCATT\_LO estimator
(\textit{top panel}), and the number of change points (NBBLOCKS) from
the Bayesian blocks algorithm (\textit{bottom panel}). In case of
NBBLOCKS, the 3 sigma quantile is still within NBBLOCKS=1.}
\label{fig:simdists2}
\end{figure}

\begin{figure}
\centering \includegraphics[width=1\columnwidth]{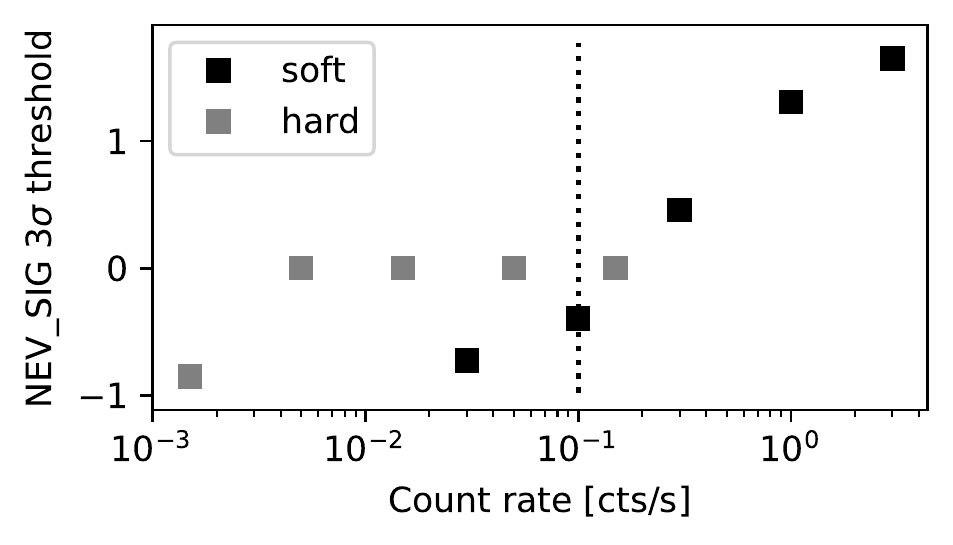}
\includegraphics[width=1\columnwidth]{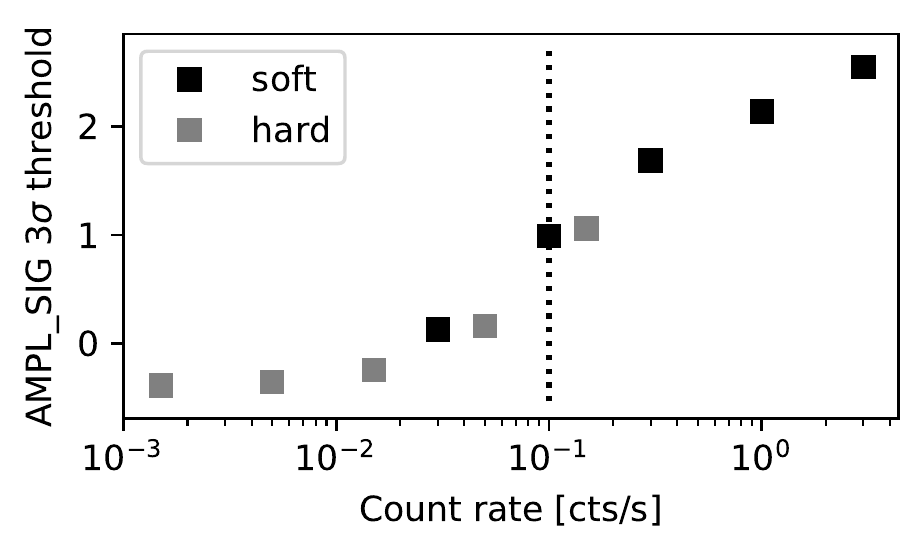} 
\caption{Calibrated thresholds as a function of count rate. Points show $3\sigma$
extreme for each simulation, for excess variance (top), amplitude
maximum deviation (middle panel) and Bayesian excess variance (bottom
panel). The vertical dotted line indicates the typical uncertainty
on $\bar{R}_{S}$.}
\label{fig:simthreshs}
\end{figure}

\section{Results}

\subsection{Thresholds for low false positive rates\label{subsec:Thresholds-for-low}}

\label{sec:calibratepurity} To find a reliable threshold corresponding
to a low false positive rate, the \textit{constant} data set is used.
The idea is to choose a threshold that rarely triggers in this non-variable
data set and that it corresponds to some desired p-value. For a given
method, its variability estimator is computed for each simulated \textit{constant}
light curve. This gives an estimator distribution for non-variable
sources. Figures~\ref{fig:simdists1} and \ref{fig:simdists2} show
the distributions for the estimators of the maximum amplitude, excess
variance, bexvar and Bayesian blocks methods at various input count
rates. The expectation is that given that these are non-variable sources,
a significance value as extreme as $3\sigma$ indeed occurs with a
frequency (p-value) corresponding to 0.27\%. However, Figure~\ref{fig:simdists1}
illustrates that the distribution of significance estimators (x-axis,
in units of $\sigma$), does not exactly match the observed $2\sigma$,
$3\sigma$, $4\sigma$ quantiles of the distribution. This is because
the significance estimators employ approximations such as Gaussian
errors. For example, the normalised excess variance under-estimates
the significance: 1$\sigma$ significances almost never occur in data
sets with $<1\mathrm{cts/s}$. The amplitude maximum deviation also
appears to slightly underestimate the significance (AMPL\_SIG>2 is
reached in fewer than 1\% of cases). The deviations are most extreme
in the low count rate regime, where AMPL\_SIG and NEV\_SIG values
never exceed 1 by chance.

We choose the threshold at the $3\sigma$ equivalent quantile of the
distributions from Figures~\ref{fig:simdists1} and \ref{fig:simdists2}.
This corresponds to a 0.3\% false positive rate at that count rate.
This approach can be applied to any estimators, whether it indicates
a significance (like AMPL\_SIG and NEV\_SIG) or an effect size (like
SCATT\_LO and NBBLOCKS). Figure~\ref{fig:simthreshs} shows these
thresholds as a function of count rate. For excess variance (top panel
of Figure~\ref{fig:simthreshs}), and AMPL\_SIG (middle panel) it
lies in the $0-2\sigma$ range, and decreases towards low count rates.
Recall that AMPL\_MAX measures the distance between the lower error
bar of the highest point and the upper error bar of the lowest point.
When counts are low, the conservatively estimated error bars are large
and mostly overlapping, giving very small or negative AMPL\_MAX values.
The significance further judges the distance by the error bars. This
leads to AMPL\_SIG decreasing with count rate and to low numbers.
A similar effect occurs with NEV\_SIG due to the overly conservative
error bars (see~§\ref{sec:srcestimates}). 

\begin{table}
\caption{Reliable thresholds. Thresholds are calibrated to a $3\sigma$ false
positive rate at all count rates. The expected number of false positives
is derived assuming a count rate distribution.}
\label{tab:threshfuncs}

\centering

\begin{tabular}{ccc}
Method & Threshold & Expected false positives\tabularnewline
\hline 
\hline 
AMPL\_SIG & \threshworstAMPLSIGbandsoft & \falseposnumAMPLSIGbandsoft\tabularnewline
NEV\_SIG & \threshworstNEVSIGbandsoft & \falseposnumNEVSIGbandsoft\tabularnewline
FVAR\_SIG & \threshworstFVARSIGbandsoft & \falseposnumFVARSIGbandsoft\tabularnewline
SCATT\_LO & \threshworstSCATTLObandsoft & \falseposnumSCATTLObandsoft\tabularnewline
\end{tabular}
\end{table}

For the Bayesian excess variance, the SCATT\_LO threshold has a peak
and does not rise towards the extreme count rates (bottom panel of
Figure~\ref{fig:simthreshs}). The difference to AMPL\_SIG and NEV\_SIG
may be because SCATT\_LO measures an effect size, not a significance.
For Bayesian blocks, the significance threshold (bottom panel of Figure~\ref{fig:simdists2})
is always at $n_{\mathrm{cp}}=1$. That is, when the Bayesian block
splits the light curve, it is reliable.

To derive a significance threshold for use in practice, a count rate
distribution has to be assumed. Taking all simulations together would
imply a log-uniform count rate distribution. In reality, the source
count rates are peaking between $0.1$ and 1 cts/s and decline towards
the high end approximately like a powerlaw with index $-1.5$. To
be conservative, we choose the highest threshold across the simulated
count rates, and present them in Table~\ref{tab:threshfuncs}. Because
the sample is dominated by faint sources, this leads to a lower false
positive rate than $3\sigma$ (0.3\%), and thus fewer than the naively
expected number of false positives (75/27910). We estimate the expected
false positive rate by weighing the simulations by a count rate powerlaw
$\bar{R}_{S}^{-1.5}$. Table~\ref{tab:threshfuncs} lists the expected
number of false positives in eFEDS for each method. For the Bayesian
excess variance, the expected number is a dozen, for the other methods
essentially no outliers are expected.

With the significance threshold chosen, we can now test which method
is most sensitive to detect variability.

\subsection{Sensitivity evaluation\label{subsec:Sensitivity-evaluation}}

\begin{figure*}
\centering \includegraphics[width=1\textwidth]{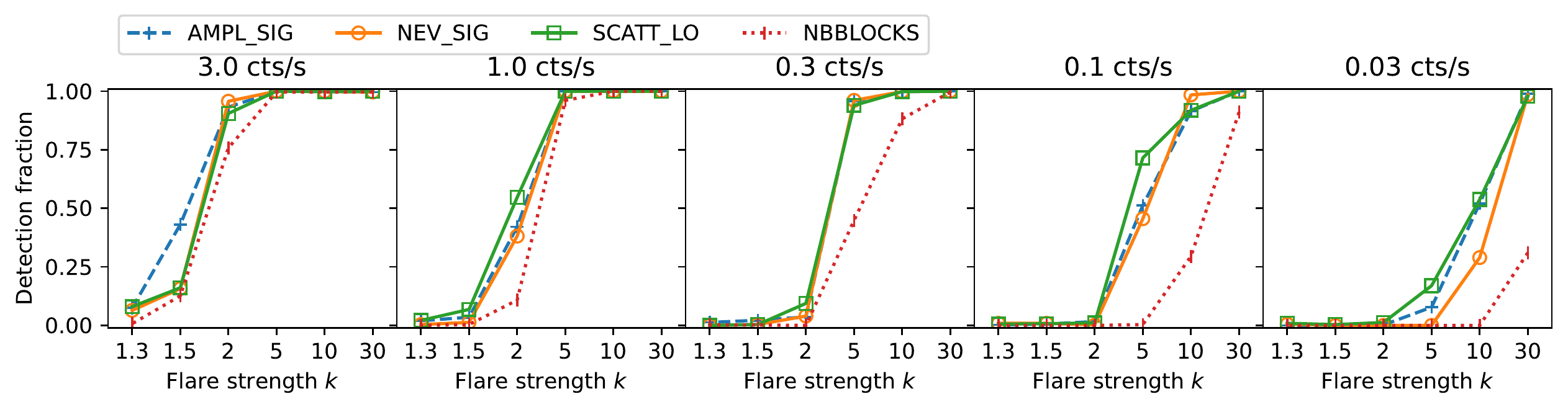}
\caption{Sensitivity of methods to detecting flares. Panels represent simulations
with increasing input count rates from left to right. Flares of varying
strengths are injected (x-axis). The fraction of objects where the
method gives an estimate above the significance threshold is shown
in the y-axis. At very high counts (left panels), the AMPL\_SIG has
the highest fraction. At medium and low counts, SCATT\_LO has the
highest detection fraction overall.}
\label{fig:simsensflare}
\end{figure*}

\begin{figure*}
\centering \includegraphics[width=1\textwidth]{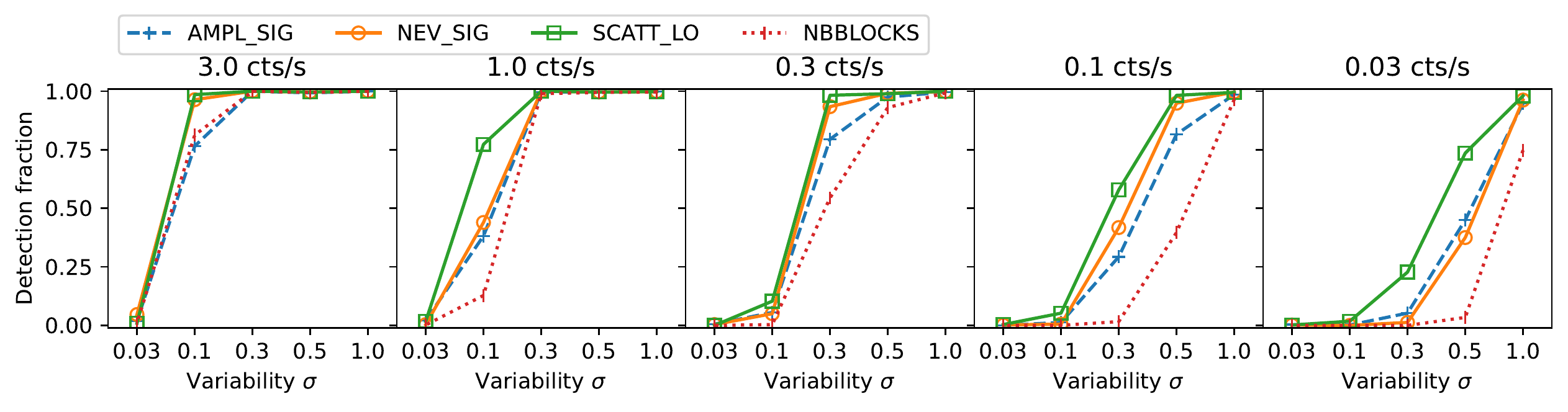}
\caption{As in Fig.~\ref{fig:simsensflare}, but for simulated white log-normal
variability of varying strength $\sigma$ (in dex). SCATT\_LO has
the highest detection fraction overall.}
\label{fig:simsensgaussvar}
\end{figure*}
\begin{figure*}
\centering \includegraphics[width=1\textwidth]{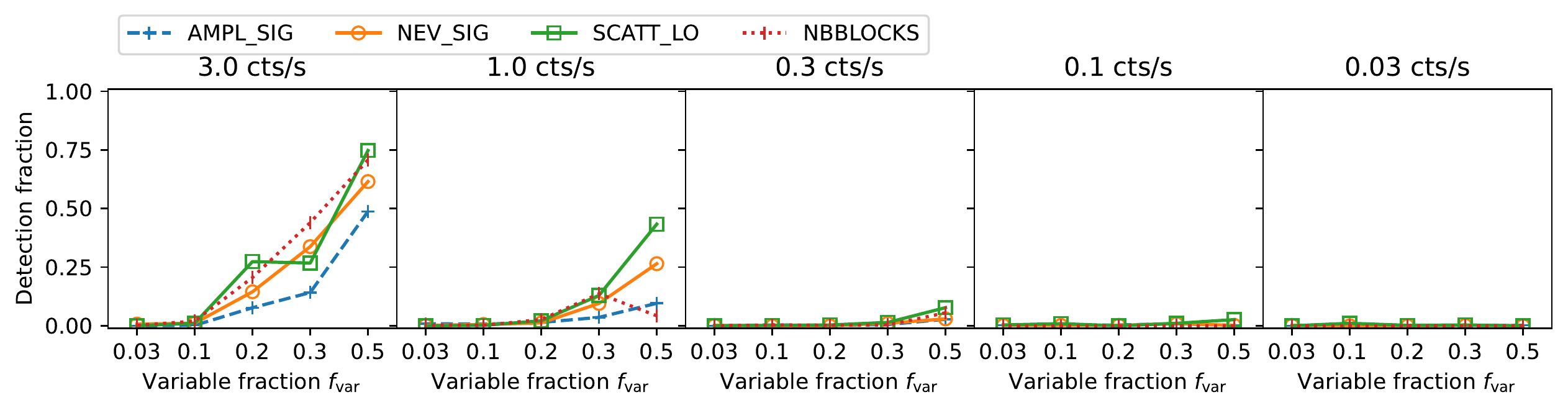}
\caption{As in Fig.~\ref{fig:simsensflare}, but for simulated red-noise variability
of varying fraction $f_{\mathrm{var}}$. Bayesian blocks has the highest
detection fraction across all panels.}
\label{fig:simsensredvar}
\end{figure*}

The goal of this section is to identify the right variability method
for detecting each type of variability. While the different methods
are based on the same data (binned light curves), they vary in assumptions
and how they use this information. Some ignore the order, some ignore
all but the most extreme points.

We quantify the sensitivity of each method using the \textit{gaussvar},
\textit{redvar} and \textit{flare} data sets. To do so, we apply the
methods to each simulated light curve and compute the fraction above
the significance thresholds calibrated in the previous section. This
fraction is the completeness of the method. The simulations vary input
count rate, strength and type of variability, allowing an in-depth
look at the behaviour of the different methods. This allows us to
characterise the detection efficiency by type (Figure~\ref{fig:simsensflare}\textit{\emph{
for}} \textit{flare}, Figure~\ref{fig:simsensgaussvar}\textit{\emph{
for }}\textit{gaussvar}, Figure~\ref{fig:simsensredvar}\textit{\emph{
for }}\textit{redvar}), but also down to which $k$ and $\sigma$
values variability the methods are sensitive.

For flares (Figure~\ref{fig:simsensflare}), amplitude maximum deviation
and Bayesian excess variance are the most sensitive method. The amplitude
maximum deviation performs better at very high count rates, while
the Bayesian excess variance is most complete in all other situations.
Flares of a factor of 5-10 are detectable for typical eROSITA sources
with these methods. The normalised excess variance has comparable
completeness as the Bayesian excess variance, except at the lowest
count rates. Bayesian blocks is less efficient at all count rates.

For white noise source variability from a log-normal distribution
(Figure~\ref{fig:simsensgaussvar}), the Bayesian excess variance
is the most sensitive method at all count rates, followed by the normalised
excess variance, amplitude maximum deviation and Bayesian blocks.
In the more realistic red noise scenario with a small variable fraction
(Figure~\ref{fig:simsensredvar}), Bayesian excess variance also
performs best in all but one simulation subgroup. Here, however, the
Bayesian block algorithm performs similarly well. Overall, only large
fractional variances ($f_{\mathrm{var}}\geq30\%$) in the high count-rate
sources ($\bar{R}_{S}>1\mathrm{cts/s}$) can be detected. To compare
Figure~\ref{fig:simsensgaussvar} and \ref{fig:simsensredvar}, $\sigma\approx f_{\mathrm{var}}/2$,
if the random walk is well sampled.

\section{Discussion and Conclusion\label{sec:Discussion-and-Conclusion}}

This work focused on characterizing four methods for detecting source
variable X-ray sources: the amplitude maximum deviation and Bayesian
excess variance, normalised excess variance and Bayesian blocks. 

\subsection{Bexvar}

The Bayesian excess variance (bexvar) is presented here for the first
time. It is a fully Poissonian way to quantify source variability
in the presence of background. We publish the bexvar code as free
and open source Python software at \url{https://gitlab.mpcdf.mpg.de/jbuchner/bexvar}.

Currently, a simple time-independent log-normal distribution is assumed.
However, the hierarchical Bayesian model is extensible. More complex
variability models, such as fitting linear, exponentially declining,
or periodic (sinosoidal) signals and potentially auto-regressive moving
average processes \citep[see e.g.,][]{Kelly2014} can be implemented
and applied to Poisson data.

Employing the Bayesian excess variance as a method to detect variability
has some limitations. Requiring the 10\% quantile on the log-normal
scatter to exceed \threshworstSCATTLObandsoft~dex makes a cut on
significance \emph{and} effect size. This will not detect barely variable
sources even when the data are excellent. Bayesian model comparison
of a constant model to a log-normal model may be even more powerful
discriminator. Indeed, the strength of the Bayesian excess variance
is not in the detection of variability, but in variability quantification.
In appendix~\ref{sec:Parameter-Recovery}, we verify that the input
parameters can be accurately and reliably retrieved.

\subsection{Efficient detection of variable sources for eROSITA}

When comparing the four methods, we find that each method has strengths
in detecting certain types of variability. For flares, amplitude maximum
deviation is both sensitive and simple to compute. It is optimized
to detect single outliers, so it is not surprising that it performs
well here. However, it is perhaps somewhat surprising that Bayesian
excess variance performs similarly well. This may be because it models
the Poisson variations carefully, and is sensitive to excess variance.
Both methods outperform the normalised excess variance and Bayesian
blocks. We presume that carefully modelling the Poisson (source and
background) noise leads to Bayesian excess variance outperforming
the classical normalised excess variance.

For intrinsic log-normal variability, the Bayesian excess variance
performs best overall. Comparing across panels in Figure~\ref{fig:simsensgaussvar},
it allows detecting variability in sources three times fainter than
Bayesian blocks with Gaussian noise. This is expected, because it
models the chosen simulated white noise process. At low count rates,
it substantially outperforms the normalised excess variance, which
assumes the same model but uses Gaussian approximations. It is surprising
that amplitude maximum deviation also outperforms the normalised excess
variance, even though the latter considers all points. However, the
trends change when white noise is replaced with a more realistic red
noise. In that case, because data points are correlated in time, the
order becomes important. Bayesian blocks, the only method tested here
that takes the order of data points into account, performs better
in this case. However, the detection efficiencies for realistic source
parameters are very modest for all methods.

For observing patterns yielding only few ($<20$) light curve data
points, amplitude maximum deviation and Bayesian blocks are quick
but effective methods, and therefore recommended for large surveys.
The Bayesian excess variance requires more computational resources,
but identifies a larger number of variable sources, especially in
the low-count regime. This is demonstrated by our simulations but
also true in practice. All four presented methods are applied to the
eFEDS observations in \citet{2021arXiv210614523B}, at the same false
positive rate (0.3\%). The 65 sources significantly detected by one
of the four methods, primarily consist of flaring stars and variable
active galactic nuclei. All methods were able to detect variability
among the 2\% brightest sources of the eFEDS sample. However, the
Bayesian excess variance more than doubled the number of sources,
and detects variability down to source count rates which encompass
20\% of the eFEDS sample.

\subsection{Outlook for the eROSITA all-sky survey}

\begin{figure}
\includegraphics[width=1\columnwidth]{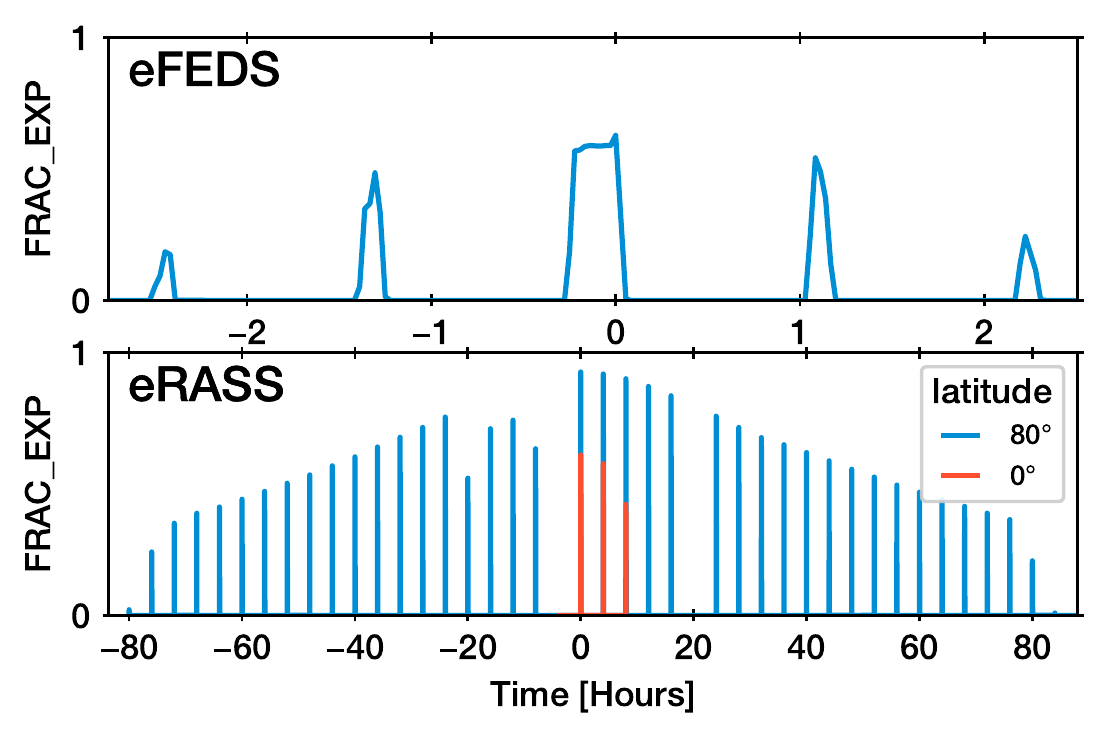}
\caption{Fractional exposure for an arbitrary eFEDS source (top panel) and
two eRASS source at different ecliptic latitudes (bottom panel).}
\label{fig:fracexp_efeds_erass}
\end{figure}

In some regards, the eFEDS survey investigated here has similar properties
to the \emph{eROSITA} all-sky survey (eRASS). eRASS will ultimately
consist of eight all-sky scan. These scans take six months to complete.
With exception of sources near the ecliptic poles, which require different
treatment (blue in Figure~\ref{fig:fracexp_efeds_erass}), most sources
are visited over a period of a few days, and covered repeatedly for
a few minutes (orange in Figure~\ref{fig:fracexp_efeds_erass}).
This cadence pattern (3-8 chunks of observations, each resolved into
multiple time bins) is not unlike the eFEDS light curve cadence. The
total exposure time of eFEDS is designed to be comparable to that
of eRASS. Therefore, sources of similar count distributions are expected.
Thus the simulation setup to test and compare variability methods,
as well as the derived significance thresholds, have applicability
also to the final eRASS observations.

In conclusion, we recommend the Bayesian excess variance and amplitude
maximum deviation methods for the detection of variable sources in
\emph{eROSITA}, with the significance thresholds specified in Table~\ref{tab:threshfuncs}.
However, variability detection and characterization methods benefit
past, present and future high-energy experiments. Improvements in
methodology can lead to new discoveries in archival data and allow
future mission such as Athena \citep{Nandra2013} and Einstein Probe
\citep{Yuan2015} to deliver more events in real time.

\section{Software packages}

matplotlib \citep{matplotlib}, UltraNest\footnote{https://johannesbuchner.github.io/UltraNest/}
\citep{Buchner2021}, astropy\footnote{https://www.astropy.org/}
\citep{AstropyCollaboration2013,AstropyCollaboration2018}, gammapy\footnote{https://gammapy.org/}
\citep{Deil2017gammapy,Nigro2019gammapy}.
\begin{acknowledgements}
We thank the anonymous referee for insightful comments that improved
the paper. JB thanks Mirko Krumpe for comments on the manuscript. 

This work is based on data from eROSITA, the soft X-ray instrument
aboard SRG, a joint Russian-German science mission supported by the
Russian Space Agency (Roskosmos), in the interests of the Russian
Academy of Sciences represented by its Space Research Institute (IKI),
and the Deutsches Zentrum für Luft- und Raumfahrt (DLR). The SRG spacecraft
was built by Lavochkin Association (NPOL) and its subcontractors,
and is operated by NPOL with support from the Max Planck Institute
for Extraterrestrial Physics (MPE). 

The development and construction of the eROSITA X-ray instrument was
led by MPE, with contributions from the Dr. Karl Remeis Observatory
Bamberg \& ECAP (FAU Erlangen-Nuernberg), the University of Hamburg
Observatory, the Leibniz Institute for Astrophysics Potsdam (AIP),
and the Institute for Astronomy and Astrophysics of the University
of Tübingen, with the support of DLR and the Max Planck Society. The
Argelander Institute for Astronomy of the University of Bonn and the
Ludwig Maximilians Universität Munich also participated in the science
preparation for eROSITA. 

The eROSITA data shown here were processed using the eSASS/NRTA software
system developed by the German eROSITA consortium.
\end{acknowledgements}

\appendix

\section{Parameter Recovery}

\label{sec:Parameter-Recovery}

\begin{figure}
\includegraphics[width=1\columnwidth]{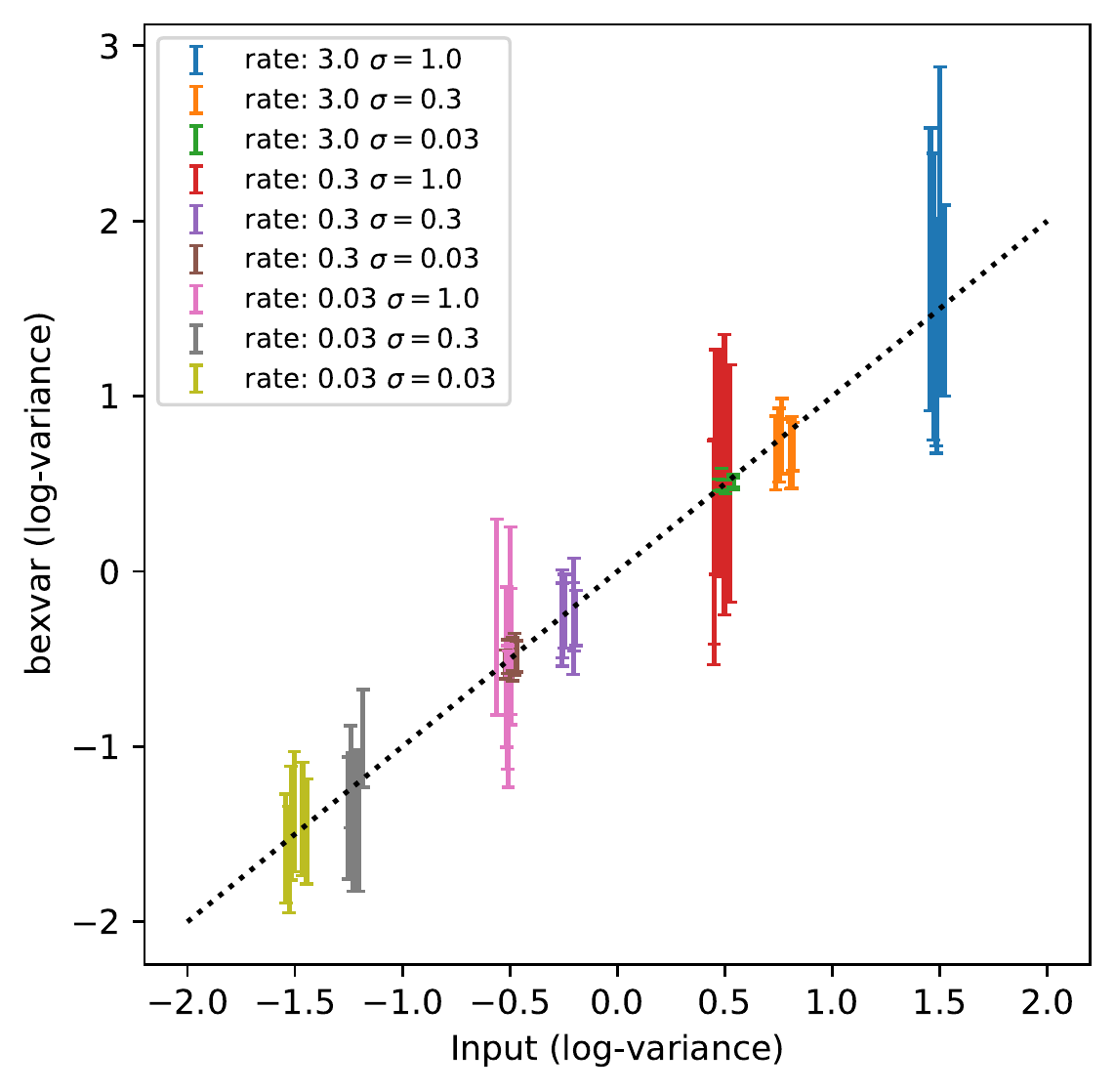}

\caption{\label{fig:recover-bexvar}Inferred Bayesian excess variance for various
simulated variances and count rates. Error bars indicate the 95\%
credible interval for an arbitrary subset of the simulations. A small
displacement in the x-axis is added to each data point for clarity.
The dotted line indicates the 1:1 correspondence.}
\end{figure}

\begin{figure}
\includegraphics[width=1\columnwidth]{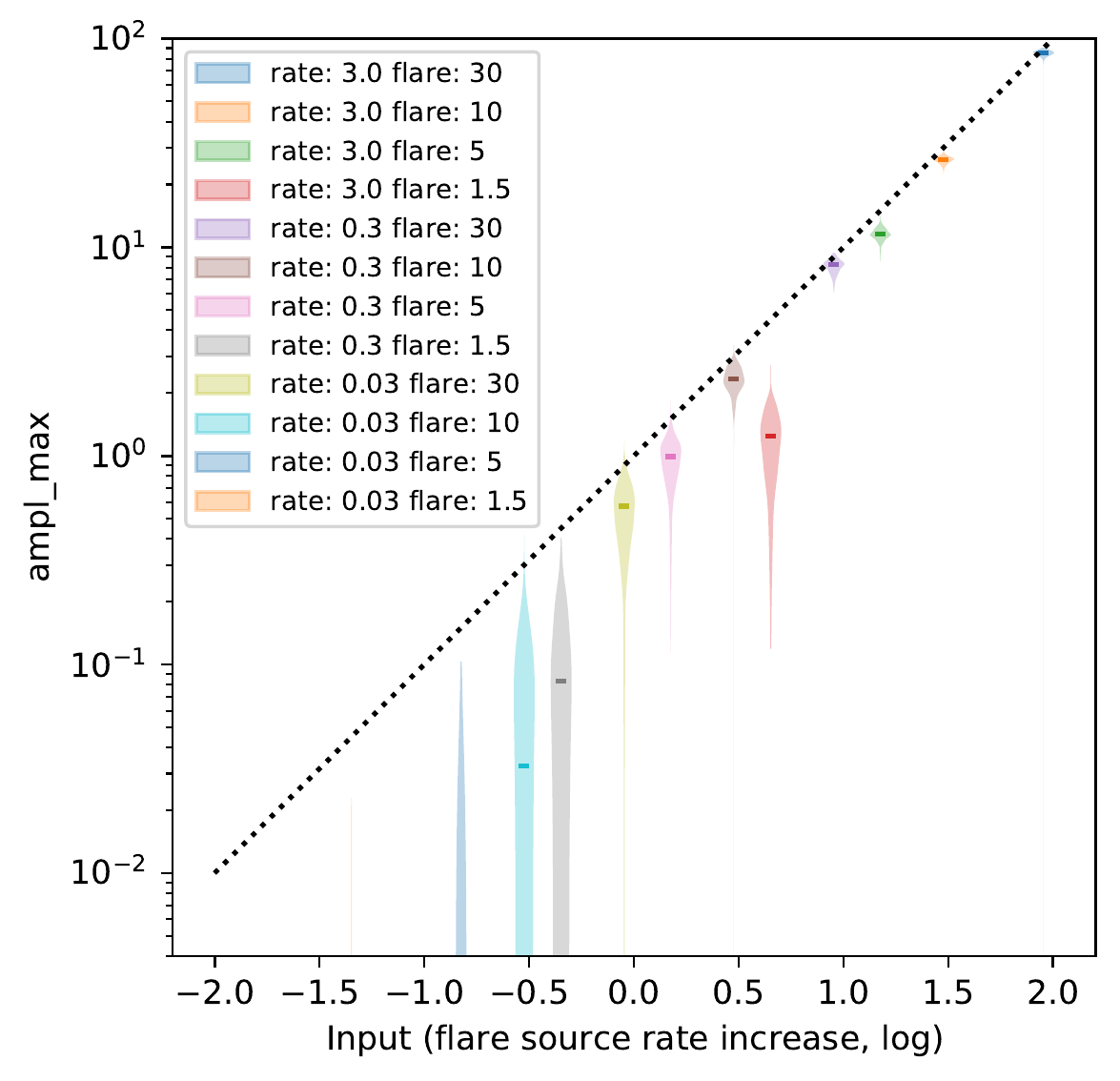}

\caption{\label{fig:recovery-ampl}Inferred amplitude maximum deviation distribution
(violin plots) for various input flare strengths. The dotted line
indicates the 1:1 correspondence. Most ampl\_max values typically
lie below this line.}
\end{figure}

While the focus of this work is on the detection of variability, some
of the methods employed quantify the variability. This depends on
the assumed variability model, which are for example step functions
in Bayesian blocks and (log) normal count rate scatter for the (Bayesian)
excess variance. Based on the flare and white noise simulations, the
recovery of methods that closely resemble these signals are investigated.

Figure~\ref{fig:recover-bexvar} compares the injected log-variance
to the inferred Bayesian excess variance in the \emph{gaussvar} simulations.
At all count rates and variability levels, the injected variance is
correctly recovered. 

Figure~\ref{fig:recovery-ampl} compares the injected flare amplitude
to the ampl\_max measure. The distribution of values is indicated
as a violin plot. Recall that ampl\_max measures the span between
error bars, and thus tends to conservatively under-estimate the flare
strength. Nevertheless, there is good overall correspondence.

\bibliographystyle{aa}
\bibliography{agn,variability}

\begin{thebibliography}{55}
\expandafter\ifx\csname natexlab\endcsname\relax\def\natexlab#1{#1}\fi

\bibitem[{{Armstrong} {et~al.}(2016){Armstrong}, {Kirk}, {Lam}, {McCormac},
  {Osborn}, {Spake}, {Walker}, {Brown}, {Kristiansen}, {Pollacco}, {West}, \&
  {Wheatley}}]{Armstrong2016}
{Armstrong}, D.~J., {Kirk}, J., {Lam}, K.~W.~F., {et~al.} 2016, \mnras, 456,
  2260

\bibitem[{{Astropy Collaboration} {et~al.}(2018){Astropy Collaboration},
  {Price-Whelan}, {Sip{\H{o}}cz}, {G{\"u}nther}, {Lim}, {Crawford}, {Conseil},
  {Shupe}, {Craig}, {Dencheva}, {Ginsburg}, {VanderPlas}, {Bradley},
  {P{\'e}rez-Su{\'a}rez}, {de Val-Borro}, {Aldcroft}, {Cruz}, {Robitaille},
  {Tollerud}, {Ardelean}, {Babej}, {Bach}, {Bachetti}, {Bakanov}, {Bamford},
  {Barentsen}, {Barmby}, {Baumbach}, {Berry}, {Biscani}, {Boquien}, {Bostroem},
  {Bouma}, {Brammer}, {Bray}, {Breytenbach}, {Buddelmeijer}, {Burke},
  {Calderone}, {Cano Rodr{\'\i}guez}, {Cara}, {Cardoso}, {Cheedella}, {Copin},
  {Corrales}, {Crichton}, {D'Avella}, {Deil}, {Depagne}, {Dietrich}, {Donath},
  {Droettboom}, {Earl}, {Erben}, {Fabbro}, {Ferreira}, {Finethy}, {Fox},
  {Garrison}, {Gibbons}, {Goldstein}, {Gommers}, {Greco}, {Greenfield},
  {Groener}, {Grollier}, {Hagen}, {Hirst}, {Homeier}, {Horton}, {Hosseinzadeh},
  {Hu}, {Hunkeler}, {Ivezi{\'c}}, {Jain}, {Jenness}, {Kanarek}, {Kendrew},
  {Kern}, {Kerzendorf}, {Khvalko}, {King}, {Kirkby}, {Kulkarni}, {Kumar},
  {Lee}, {Lenz}, {Littlefair}, {Ma}, {Macleod}, {Mastropietro}, {McCully},
  {Montagnac}, {Morris}, {Mueller}, {Mumford}, {Muna}, {Murphy}, {Nelson},
  {Nguyen}, {Ninan}, {N{\"o}the}, {Ogaz}, {Oh}, {Parejko}, {Parley}, {Pascual},
  {Patil}, {Patil}, {Plunkett}, {Prochaska}, {Rastogi}, {Reddy Janga},
  {Sabater}, {Sakurikar}, {Seifert}, {Sherbert}, {Sherwood-Taylor}, {Shih},
  {Sick}, {Silbiger}, {Singanamalla}, {Singer}, {Sladen}, {Sooley},
  {Sornarajah}, {Streicher}, {Teuben}, {Thomas}, {Tremblay}, {Turner},
  {Terr{\'o}n}, {van Kerkwijk}, {de la Vega}, {Watkins}, {Weaver}, {Whitmore},
  {Woillez}, {Zabalza}, \& {Astropy Contributors}}]{AstropyCollaboration2018}
{Astropy Collaboration}, {Price-Whelan}, A.~M., {Sip{\H{o}}cz}, B.~M., {et~al.}
  2018, \aj, 156, 123

\bibitem[{{Astropy Collaboration} {et~al.}(2013){Astropy Collaboration},
  {Robitaille}, {Tollerud}, {Greenfield}, {Droettboom}, {Bray}, {Aldcroft},
  {Davis}, {Ginsburg}, {Price-Whelan}, {Kerzendorf}, {Conley}, {Crighton},
  {Barbary}, {Muna}, {Ferguson}, {Grollier}, {Parikh}, {Nair}, {Unther},
  {Deil}, {Woillez}, {Conseil}, {Kramer}, {Turner}, {Singer}, {Fox}, {Weaver},
  {Zabalza}, {Edwards}, {Azalee Bostroem}, {Burke}, {Casey}, {Crawford},
  {Dencheva}, {Ely}, {Jenness}, {Labrie}, {Lim}, {Pierfederici}, {Pontzen},
  {Ptak}, {Refsdal}, {Servillat}, \& {Streicher}}]{AstropyCollaboration2013}
{Astropy Collaboration}, {Robitaille}, T.~P., {Tollerud}, E.~J., {et~al.} 2013,
  \aap, 558, A33

\bibitem[{{Bachetti} {et~al.}(2014){Bachetti}, {Harrison}, {Walton},
  {Grefenstette}, {Chakrabarty}, {F{\"u}rst}, {Barret}, {Beloborodov}, {Boggs},
  {Christensen}, {Craig}, {Fabian}, {Hailey}, {Hornschemeier}, {Kaspi},
  {Kulkarni}, {Maccarone}, {Miller}, {Rana}, {Stern}, {Tendulkar}, {Tomsick},
  {Webb}, \& {Zhang}}]{Bachetti2014}
{Bachetti}, M., {Harrison}, F.~A., {Walton}, D.~J., {et~al.} 2014, \nat, 514,
  202

\bibitem[{{Barlow}(2003)}]{Barlow2003}
{Barlow}, R. 2003, in Statistical Problems in Particle Physics, Astrophysics,
  and Cosmology, ed. L.~{Lyons}, R.~{Mount}, \& R.~{Reitmeyer}, 250

\bibitem[{{Boller} {et~al.}(2016){Boller}, {Freyberg}, {Tr{\"u}mper}, {Haberl},
  {Voges}, \& {Nandra}}]{Boller2016}
{Boller}, T., {Freyberg}, M.~J., {Tr{\"u}mper}, J., {et~al.} 2016, \aap, 588,
  A103

\bibitem[{{Boller} {et~al.}(2021){Boller}, {Schmitt}, {Buchner}, {Freyberg},
  {Georgakakis}, {Liu}, {Robrade}, {Merloni}, {Nandra}, {Malyali}, {Krumpe},
  {Salvato}, \& {Dwelly}}]{2021arXiv210614523B}
{Boller}, T., {Schmitt}, J.~H.~M.~M., {Buchner}, J., {et~al.} 2021, arXiv
  e-prints, arXiv:2106.14523

\bibitem[{{Brunner} {et~al.}(2021){Brunner}, {Liu}, {Lamer}, {Georgakakis},
  {Merloni}, {Brusa}, {Bulbul}, {Dennerl}, {Friedrich}, {Liu}, {Maitra},
  {Nandra}, {Ramos-Ceja}, {Sanders}, {Stewart}, {Boller}, {Buchner}, {Clerc},
  {Comparat}, {Dwelly}, {Eckert}, {Finoguenov}, {Freyberg}, {Ghirardini},
  {Gueguen}, {Haberl}, {Kreykenbohm}, {Krumpe}, {Osterhage}, {Pacaud},
  {Predehl}, {Reiprich}, {Robrade}, {Salvato}, {Santangelo}, {Schrabback},
  {Schwope}, \& {Wilms}}]{Brunner2021}
{Brunner}, H., {Liu}, T., {Lamer}, G., {et~al.} 2021, arXiv e-prints,
  arXiv:2106.14517

\bibitem[{Buchner(2016)}]{buchner2016statistical}
Buchner, J. 2016, Statistics and Computing, 26, 383

\bibitem[{{Buchner}(2019)}]{Buchner2019c}
{Buchner}, J. 2019, \pasp, 131, 108005

\bibitem[{{Buchner}(2021)}]{Buchner2021}
{Buchner}, J. 2021, The Journal of Open Source Software, 6, 3001

\bibitem[{{Cameron}(2011)}]{Cameron2011}
{Cameron}, E. 2011, \pasa, 28, 128

\bibitem[{{De Luca} {et~al.}(2021){De Luca}, {Salvaterra}, {Belfiore},
  {Carpano}, {D'Agostino}, {Haberl}, {Israel}, {Law-Green}, {Lisini},
  {Marelli}, {Novara}, {Read}, {Rodriguez-Castillo}, {Rosen}, {Salvetti},
  {Tiengo}, {Vianello}, {Watson}, {Delvaux}, {Dickens}, {Esposito}, {Greiner},
  {H{\"a}mmerle}, {Kreikenbohm}, {Kreykenbohm}, {Oertel}, {Pizzocaro}, {Pye},
  {Sandrelli}, {Stelzer}, {Wilms}, \& {Zagaria}}]{DeLuca2021}
{De Luca}, A., {Salvaterra}, R., {Belfiore}, A., {et~al.} 2021, \aap, 650, A167

\bibitem[{{Debosscher} {et~al.}(2007){Debosscher}, {Sarro}, {Aerts}, {Cuypers},
  {Vandenbussche}, {Garrido}, \& {Solano}}]{Debosscher2007}
{Debosscher}, J., {Sarro}, L.~M., {Aerts}, C., {et~al.} 2007, \aap, 475, 1159

\bibitem[{{Deil} {et~al.}(2017){Deil}, {Zanin}, {Lefaucheur}, {Boisson},
  {Khelifi}, {Terrier}, {Wood}, {Mohrmann}, {Chakraborty}, {Watson},
  {Lopez-Coto}, {Klepser}, {Cerruti}, {Lenain}, {Acero}, {Djannati-Ata{\"\i}},
  {Pita}, {Bosnjak}, {Trichard}, {Vuillaume}, {Donath}, {Consortium}, {King},
  {Jouvin}, {Owen}, {Sipocz}, {Lennarz}, {Voruganti}, {Spir-Jacob}, {Ruiz}, \&
  {Arribas}}]{Deil2017gammapy}
{Deil}, C., {Zanin}, R., {Lefaucheur}, J., {et~al.} 2017, in International
  Cosmic Ray Conference, Vol. 301, 35th International Cosmic Ray Conference
  (ICRC2017), 766

\bibitem[{{Ducci} {et~al.}(2020){Ducci}, {Ji}, {Haberl}, {Rau}, {Sasaki},
  {Wilms}, {Kreykenbohm}, {Weber}, {Werner}, {Santangelo}, {Maitra}, {Carpano},
  {Buchner}, {Schwope}, {Suleimanov}, \& {Doroshenko}}]{Ducci2020}
{Ducci}, L., {Ji}, L., {Haberl}, F., {et~al.} 2020, {SRG/eROSITA discovery of a
  bright supersoft X-ray emission from the classical nova AT 2018bej in the
  Large Magellanic Cloud}

\bibitem[{{Edelson} {et~al.}(2002){Edelson}, {Turner}, {Pounds}, {Vaughan},
  {Markowitz}, {Marshall}, {Dobbie}, \& {Warwick}}]{Edelson2002}
{Edelson}, R., {Turner}, T.~J., {Pounds}, K., {et~al.} 2002, \apj, 568, 610

\bibitem[{{Edelson} {et~al.}(1990){Edelson}, {Krolik}, \& {Pike}}]{Edelson1990}
{Edelson}, R.~A., {Krolik}, J.~H., \& {Pike}, G.~F. 1990, \apj, 359, 86

\bibitem[{{Gehrels}(1986)}]{Gehrels1986}
{Gehrels}, N. 1986, \apj, 303, 336

\bibitem[{{Gehrels} {et~al.}(2004){Gehrels}, {Chincarini}, {Giommi}, {Mason},
  {Nousek}, {Wells}, {White}, {Barthelmy}, {Burrows}, {Cominsky}, {Hurley},
  {Marshall}, {M{\'e}sz{\'a}ros}, {Roming}, {Angelini}, {Barbier}, {Belloni},
  {Campana}, {Caraveo}, {Chester}, {Citterio}, {Cline}, {Cropper}, {Cummings},
  {Dean}, {Feigelson}, {Fenimore}, {Frail}, {Fruchter}, {Garmire}, {Gendreau},
  {Ghisellini}, {Greiner}, {Hill}, {Hunsberger}, {Krimm}, {Kulkarni}, {Kumar},
  {Lebrun}, {Lloyd-Ronning}, {Markwardt}, {Mattson}, {Mushotzky}, {Norris},
  {Osborne}, {Paczynski}, {Palmer}, {Park}, {Parsons}, {Paul}, {Rees},
  {Reynolds}, {Rhoads}, {Sasseen}, {Schaefer}, {Short}, {Smale}, {Smith},
  {Stella}, {Tagliaferri}, {Takahashi}, {Tashiro}, {Townsley}, {Tueller},
  {Turner}, {Vietri}, {Voges}, {Ward}, {Willingale}, {Zerbi}, \&
  {Zhang}}]{Gehrels2004}
{Gehrels}, N., {Chincarini}, G., {Giommi}, P., {et~al.} 2004, \apj, 611, 1005

\bibitem[{{Gehrels} \& {M{\'e}sz{\'a}ros}(2012)}]{Gehrels2012}
{Gehrels}, N. \& {M{\'e}sz{\'a}ros}, P. 2012, Science, 337, 932

\bibitem[{{Heil} {et~al.}(2015){Heil}, {Uttley}, \& {Klein-Wolt}}]{Heil2015}
{Heil}, L.~M., {Uttley}, P., \& {Klein-Wolt}, M. 2015, \mnras, 448, 3348

\bibitem[{{Heinze} {et~al.}(2018){Heinze}, {Tonry}, {Denneau}, {Flewelling},
  {Stalder}, {Rest}, {Smith}, {Smartt}, \& {Weiland}}]{Heinze2018}
{Heinze}, A.~N., {Tonry}, J.~L., {Denneau}, L., {et~al.} 2018, \aj, 156, 241

\bibitem[{{Holl} {et~al.}(2018){Holl}, {Audard}, {Nienartowicz}, {Jevardat de
  Fombelle}, {Marchal}, {Mowlavi}, {Clementini}, {De Ridder}, {Evans}, {Guy},
  {Lanzafame}, {Lebzelter}, {Rimoldini}, {Roelens}, {Zucker}, {Distefano},
  {Garofalo}, {Lecoeur-Ta{\"\i}bi}, {Lopez}, {Molinaro}, {Muraveva}, {Panahi},
  {Regibo}, {Ripepi}, {Sarro}, {Aerts}, {Anderson}, {Charnas}, {Barblan},
  {Blanco-Cuaresma}, {Busso}, {Cuypers}, {De Angeli}, {Glass}, {Grenon},
  {Juh{\'a}sz}, {Kochoska}, {Koubsky}, {Lanza}, {Leccia}, {Lorenz}, {Marconi},
  {Marschalk{\'o}}, {Mazeh}, {Messina}, {Mignard}, {Moitinho}, {Moln{\'a}r},
  {Morgenthaler}, {Musella}, {Ordenovic}, {Ord{\'o}{\~n}ez}, {Pagano},
  {Palaversa}, {Pawlak}, {Plachy}, {Pr{\v{s}}a}, {Riello}, {S{\"u}veges},
  {Szabados}, {Szegedi-Elek}, {Votruba}, \& {Eyer}}]{Holl2018}
{Holl}, B., {Audard}, M., {Nienartowicz}, K., {et~al.} 2018, \aap, 618, A30

\bibitem[{Hunter(2007)}]{matplotlib}
Hunter, J.~D. 2007, Computing In Science \& Engineering, 9, 90

\bibitem[{{Jayasinghe} {et~al.}(2019){Jayasinghe}, {Stanek}, {Kochanek},
  {Shappee}, {Holoien}, {Thompson}, {Prieto}, {Dong}, {Pawlak}, {Pejcha},
  {Shields}, {Pojmanski}, {Otero}, {Britt}, \& {Will}}]{Jayasinghe2019}
{Jayasinghe}, T., {Stanek}, K.~Z., {Kochanek}, C.~S., {et~al.} 2019, \mnras,
  486, 1907

\bibitem[{{Kelly} {et~al.}(2014){Kelly}, {Becker}, {Sobolewska},
  {Siemiginowska}, \& {Uttley}}]{Kelly2014}
{Kelly}, B.~C., {Becker}, A.~C., {Sobolewska}, M., {Siemiginowska}, A., \&
  {Uttley}, P. 2014, \apj, 788, 33

\bibitem[{{Kim} {et~al.}(2011){Kim}, {Protopapas}, {Byun}, {Alcock}, {Khardon},
  \& {Trichas}}]{Kim2011}
{Kim}, D.-W., {Protopapas}, P., {Byun}, Y.-I., {et~al.} 2011, \apj, 735, 68

\bibitem[{{Klebesadel} {et~al.}(1973){Klebesadel}, {Strong}, \&
  {Olson}}]{Klebesadel1973}
{Klebesadel}, R.~W., {Strong}, I.~B., \& {Olson}, R.~A. 1973, \apjl, 182, L85

\bibitem[{{Knoetig}(2014)}]{Knoetig2014}
{Knoetig}, M.~L. 2014, \apj, 790, 106

\bibitem[{{Koenig} {et~al.}(2020){Koenig}, {Wilms}, {Kreykenbohm}, {Weber},
  {Bogensberger}, {Rau}, {Merloni}, {Maitra}, {Carpano}, \& {Ji}}]{Koenig2020}
{Koenig}, O., {Wilms}, J., {Kreykenbohm}, I., {et~al.} 2020, {SRG/eROSITA
  detection of a bright flare of the millisecond pulsar PSR J1023+0038}

\bibitem[{{Kraft} {et~al.}(1991){Kraft}, {Burrows}, \& {Nousek}}]{Kraft1991}
{Kraft}, R.~P., {Burrows}, D.~N., \& {Nousek}, J.~A. 1991, \apj, 374, 344

\bibitem[{{Liu} {et~al.}(2013){Liu}, {Bregman}, {Bai}, {Justham}, \&
  {Crowther}}]{Liu2013}
{Liu}, J.-F., {Bregman}, J.~N., {Bai}, Y., {Justham}, S., \& {Crowther}, P.
  2013, \nat, 503, 500

\bibitem[{{Liu} {et~al.}(2021){Liu}, {Buchner}, {Nandra}, {Merloni}, {Dwelly},
  {Sanders}, {Salvato}, {Arcodia}, {Brusa}, {Wolf}, {Georgakakis}, {Boller},
  {Krumpe}, {Lamer}, {Waddell}, {Urrutia}, {Schwope}, {Robrade}, {Wilms},
  {Dauser}, {Comparat}, {Toba}, {Ichikawa}, {Iwasawa}, {Shen}, \& {Ibarra
  Medel}}]{2021arXiv210614522L}
{Liu}, T., {Buchner}, J., {Nandra}, K., {et~al.} 2021, arXiv e-prints,
  arXiv:2106.14522

\bibitem[{{Lorimer} {et~al.}(2007){Lorimer}, {Bailes}, {McLaughlin},
  {Narkevic}, \& {Crawford}}]{Lorimer2007}
{Lorimer}, D.~R., {Bailes}, M., {McLaughlin}, M.~A., {Narkevic}, D.~J., \&
  {Crawford}, F. 2007, Science, 318, 777

\bibitem[{{Malyali} {et~al.}(2021){Malyali}, {Rau}, {Merloni}, {Nandra},
  {Buchner}, {Liu}, {Gezari}, {Sollerman}, {Shappee}, {Trakhtenbrot}, {Arcavi},
  {Ricci}, {van Velzen}, {Goobar}, {Frederick}, {Kawka}, {Tartaglia}, {Burke},
  {Hiramatsu}, {Schramm}, {van der Boom}, {Anderson}, {Miller-Jones}, {Bellm},
  {Drake}, {Duev}, {Fremling}, {Graham}, {Masci}, {Rusholme}, {Soumagnac}, \&
  {Walters}}]{Malyali2021}
{Malyali}, A., {Rau}, A., {Merloni}, A., {et~al.} 2021, \aap, 647, A9

\bibitem[{{Masci} {et~al.}(2014){Masci}, {Hoffman}, {Grillmair}, \&
  {Cutri}}]{Masci2014}
{Masci}, F.~J., {Hoffman}, D.~I., {Grillmair}, C.~J., \& {Cutri}, R.~M. 2014,
  \aj, 148, 21

\bibitem[{{Matsuoka} {et~al.}(2009){Matsuoka}, {Kawasaki}, {Ueno}, {Tomida},
  {Kohama}, {Suzuki}, {Adachi}, {Ishikawa}, {Mihara}, {Sugizaki}, {Isobe},
  {Nakagawa}, {Tsunemi}, {Miyata}, {Kawai}, {Kataoka}, {Morii}, {Yoshida},
  {Negoro}, {Nakajima}, {Ueda}, {Chujo}, {Yamaoka}, {Yamazaki}, {Nakahira},
  {You}, {Ishiwata}, {Miyoshi}, {Eguchi}, {Hiroi}, {Katayama}, \&
  {Ebisawa}}]{2009PASJ...61..999M}
{Matsuoka}, M., {Kawasaki}, K., {Ueno}, S., {et~al.} 2009, \pasj, 61, 999

\bibitem[{{Maughan} \& {Reiprich}(2019)}]{Maughan2019}
{Maughan}, B.~J. \& {Reiprich}, T.~H. 2019, The Open Journal of Astrophysics,
  2, 9

\bibitem[{{Miniutti} {et~al.}(2019){Miniutti}, {Saxton}, {Giustini},
  {Alexander}, {Fender}, {Heywood}, {Monageng}, {Coriat}, {Tzioumis}, {Read},
  {Knigge}, {Gandhi}, {Pretorius}, \& {Ag{\'\i}s-Gonz{\'a}lez}}]{Miniutti2019}
{Miniutti}, G., {Saxton}, R.~D., {Giustini}, M., {et~al.} 2019, \nat, 573, 381

\bibitem[{{Nandra} {et~al.}(2013){Nandra}, {Barret}, {Barcons}, {Fabian}, {den
  Herder}, {Piro}, {Watson}, {Adami}, {Aird}, {Afonso}, \& et~al.}]{Nandra2013}
{Nandra}, K., {Barret}, D., {Barcons}, X., {et~al.} 2013, ArXiv e-prints

\bibitem[{{Nandra} {et~al.}(1997){Nandra}, {George}, {Mushotzky}, {Turner}, \&
  {Yaqoob}}]{Nandra1997}
{Nandra}, K., {George}, I.~M., {Mushotzky}, R.~F., {Turner}, T.~J., \&
  {Yaqoob}, T. 1997, \apj, 476, 70

\bibitem[{{Nigro} {et~al.}(2019){Nigro}, {Deil}, {Zanin}, {Hassan}, {King},
  {Ruiz}, {Saha}, {Terrier}, {Br{\"u}gge}, {N{\"o}the}, {Bird}, {Lin},
  {Aleksi{\'c}}, {Boisson}, {Contreras}, {Donath}, {Jouvin}, {Kelley-Hoskins},
  {Khelifi}, {Kosack}, {Rico}, \& {Sinha}}]{Nigro2019gammapy}
{Nigro}, C., {Deil}, C., {Zanin}, R., {et~al.} 2019, \aap, 625, A10

\bibitem[{{Palaversa} {et~al.}(2013){Palaversa}, {Ivezi{\'c}}, {Eyer},
  {Ru{\v{z}}djak}, {Sudar}, {Galin}, {Kroflin}, {Mesari{\'c}}, {Munk},
  {Vrbanec}, {Bo{\v{z}}i{\'c}}, {Loebman}, {Sesar}, {Rimoldini}, {Hunt-Walker},
  {VanderPlas}, {Westman}, {Stuart}, {Becker}, {Srdo{\v{c}}}, {Wozniak}, \&
  {Oluseyi}}]{Palaversa2013}
{Palaversa}, L., {Ivezi{\'c}}, {\v{Z}}., {Eyer}, L., {et~al.} 2013, \aj, 146,
  101

\bibitem[{{Petroff} {et~al.}(2019){Petroff}, {Hessels}, \&
  {Lorimer}}]{Petroff2019}
{Petroff}, E., {Hessels}, J.~W.~T., \& {Lorimer}, D.~R. 2019, \aapr, 27, 4

\bibitem[{{Ponti} {et~al.}(2014){Ponti}, {Mu{\~n}oz-Darias}, \&
  {Fender}}]{Ponti2014}
{Ponti}, G., {Mu{\~n}oz-Darias}, T., \& {Fender}, R.~P. 2014, \mnras, 444, 1829

\bibitem[{{Predehl} {et~al.}(2021){Predehl}, {Andritschke}, {Arefiev},
  {Babyshkin}, {Batanov}, {Becker}, {B{\"o}hringer}, {Bogomolov}, {Boller},
  {Borm}, {Bornemann}, {Br{\"a}uninger}, {Br{\"u}ggen}, {Brunner}, {Brusa},
  {Bulbul}, {Buntov}, {Burwitz}, {Burkert}, {Clerc}, {Churazov}, {Coutinho},
  {Dauser}, {Dennerl}, {Doroshenko}, {Eder}, {Emberger}, {Eraerds},
  {Finoguenov}, {Freyberg}, {Friedrich}, {Friedrich}, {F{\"u}rmetz},
  {Georgakakis}, {Gilfanov}, {Granato}, {Grossberger}, {Gueguen}, {Gureev},
  {Haberl}, {H{\"a}lker}, {Hartner}, {Hasinger}, {Huber}, {Ji}, {Kienlin},
  {Kink}, {Korotkov}, {Kreykenbohm}, {Lamer}, {Lomakin}, {Lapshov}, {Liu},
  {Maitra}, {Meidinger}, {Menz}, {Merloni}, {Mernik}, {Mican}, {Mohr},
  {M{\"u}ller}, {Nandra}, {Nazarov}, {Pacaud}, {Pavlinsky}, {Perinati},
  {Pfeffermann}, {Pietschner}, {Ramos-Ceja}, {Rau}, {Reiffers}, {Reiprich},
  {Robrade}, {Salvato}, {Sanders}, {Santangelo}, {Sasaki}, {Scheuerle},
  {Schmid}, {Schmitt}, {Schwope}, {Shirshakov}, {Steinmetz}, {Stewart},
  {Str{\"u}der}, {Sunyaev}, {Tenzer}, {Tiedemann}, {Tr{\"u}mper}, {Voron},
  {Weber}, {Wilms}, \& {Yaroshenko}}]{Predehl2020}
{Predehl}, P., {Andritschke}, R., {Arefiev}, V., {et~al.} 2021, \aap, 647, A1

\bibitem[{{Scargle} {et~al.}(2013){Scargle}, {Norris}, {Jackson}, \&
  {Chiang}}]{Scargle2013}
{Scargle}, J.~D., {Norris}, J.~P., {Jackson}, B., \& {Chiang}, J. 2013, \apj,
  764, 167

\bibitem[{{Simm} {et~al.}(2016){Simm}, {Salvato}, {Saglia}, {Ponti},
  {Lanzuisi}, {Trakhtenbrot}, {Nandra}, \& {Bender}}]{Simm2016}
{Simm}, T., {Salvato}, M., {Saglia}, R., {et~al.} 2016, \aap, 585, A129

\bibitem[{{Swank}(2006)}]{Swank2006}
{Swank}, J.~H. 2006, Advances in Space Research, 38, 2959

\bibitem[{{van Roestel} {et~al.}(2021){van Roestel}, {Duev}, {Mahabal},
  {Coughlin}, {Mr{\'o}z}, {Burdge}, {Drake}, {Graham}, {Hillenbrand}, {Bellm},
  {Kupfer}, {Delacroix}, {Fremling}, {Golkhou}, {Hale}, {Laher}, {Masci},
  {Riddle}, {Rosnet}, {Rusholme}, {Smith}, {Soumagnac}, {Walters}, {Prince}, \&
  {Kulkarni}}]{vanRoestel2021}
{van Roestel}, J., {Duev}, D.~A., {Mahabal}, A.~A., {et~al.} 2021, \aj, 161,
  267

\bibitem[{{Vaughan} {et~al.}(2003){Vaughan}, {Edelson}, {Warwick}, \&
  {Uttley}}]{Vaughan2003}
{Vaughan}, S., {Edelson}, R., {Warwick}, R.~S., \& {Uttley}, P. 2003, \mnras,
  345, 1271

\bibitem[{{Weber}(2020)}]{2020GCN.26988....1W}
{Weber}, P. 2020, GRB Coordinates Network, 26988, 1

\bibitem[{{Wilms} {et~al.}(2020){Wilms}, {Kreykenbohm}, {Weber}, {Falkner},
  {Dauser}, {Knies}, {Koenig}, {Malyali}, {Rau}, {Merloni}, {Bogensberger},
  {Brunner}, {Buchner}, {Carpano}, {Freyberg}, {Haberl}, {Maitra}, {Salvato},
  {Doroshenko}, {Ducci}, {Ji}, {Schmitt}, \& {Schwope}}]{Wilms2020}
{Wilms}, J., {Kreykenbohm}, I., {Weber}, P., {et~al.} 2020, {SRG/eROSITA
  detection of the bright, transient X-ray flare SRGt J123822.3-253206}

\bibitem[{{Yuan} {et~al.}(2015){Yuan}, {Zhang}, {Feng}, {Zhang}, {Ling},
  {Zhao}, {Deng}, {Qiu}, {Osborne}, {O'Brien}, {Willingale}, {Lapington},
  {Fraser}, \& {the Einstein Probe team}}]{Yuan2015}
{Yuan}, W., {Zhang}, C., {Feng}, H., {et~al.} 2015, arXiv e-prints,
  arXiv:1506.07735

\end{thebibliography}

\end{document}